\documentclass[twocolumn]{aastex62}
\usepackage{amsmath}
\usepackage{fancyvrb}

\newcommand{\teff}{\ensuremath{T_{\mathrm{eff}}}}
\newcommand{\logg}{\ensuremath{\log{g}}}
\newcommand{\FeH}{\ensuremath{[\mathrm{Fe}/\mathrm{H}]}}
\newcommand{\alphaM}{\ensuremath{[\alpha/\mathrm{Fe}]}}
\newcommand*{\fea}[1][]{\ensuremath{\mathbf{X}_{#1}}}
\newcommand*{\lab}[1][]{\ensuremath{\mathbf{Y}_{#1}}}

\newcommand\phoenix{\textsc{phoenix}}
\newcommand\cosha{\textsc{CoSHA}}

\defcitealias{Yan2019}{MaStarDR1}

\graphicspath{{./}{figures/}}

\received{January 1, 2018}
\revised{January 7, 2018}
\accepted{\today}
\submitjournal{ApJ}

\shorttitle{Assigning parameters to MaStar}
\shortauthors{Mej\'ia-Narv\'aez et al.}

\begin{document}

\title{\cosha{}: Code for Stellar properties Heuristic Assignment -- for the MaStar stellar library}

\correspondingauthor{Alfredo Mej\'ia-Narv\'aez}
\email{amejia@astro.unam.mx}

\author{Alfredo Mej\'ia-Narv\'aez}
\affil{Instituto de Astronom\'ia, Universidad Nacional Aut\'onoma de M\'exico \\
A. P. 70-264, C.P. 04510 \\
M\'exico, D.F., M\'exico}

\author{Gustavo Bruzual}
\affiliation{Instituto de Radioastronom\'ia y Astrof\'isica, Universidad Nacional Aut\'onoma de M\'exico \\
Campus Morelia, Michoacan \\
M\'exico C.P. 58089, M\'exico}
\nocollaboration

\author{Sebastian.~F. S\'anchez}
\affiliation{Instituto de Astronom\'ia, Universidad Nacional Aut\'onoma de M\'exico \\
A. P. 70-264, C.P. 04510 \\
M\'exico, D.F., M\'exico}
\nocollaboration

\author{Leticia Carigi}
\affiliation{Instituto de Astronom\'ia, Universidad Nacional Aut\'onoma de M\'exico \\
A. P. 70-264, C.P. 04510 \\
M\'exico, D.F., M\'exico}
\nocollaboration

\author{Jorge Barrera-Ballesteros}
\affiliation{Instituto de Astronom\'ia, Universidad Nacional Aut\'onoma de M\'exico \\
A. P. 70-264, C.P. 04510 \\
M\'exico, D.F., M\'exico}
\nocollaboration

\author{Mabel Valerdi}
\affiliation{Instituto de Astronom\'ia, Universidad Nacional Aut\'onoma de M\'exico \\
A. P. 70-264, C.P. 04510 \\
M\'exico, D.F., M\'exico}
\nocollaboration

\author{Renbin Yan}
\affiliation{Department of Physics, The Chinese University of Hong Kong,\\
Shatin, N.T.,\\
Hong Kong, China}
\nocollaboration

\author{Niv Drory}
\affiliation{McDonald Observatory, Department of Astronomy, \\
University of Texas at Austin, 1 University Station, \\
Austin, TX 78712-0259, USA}
\nocollaboration



\begin{abstract}

We introduce \cosha{}: a Code for Stellar properties Heuristic Assignment. In order to estimate the stellar properties, \cosha{} implements a Gradient Tree Boosting algorithm to label each star across the parameter space (\teff{}, \logg{}, \FeH{}, and \alphaM{}). We use \cosha{} to estimate these stellar atmospheric parameters of $22\,$k unique stars in the MaNGA Stellar Library (MaStar). To quantify the reliability of our approach, we run both internal tests using the G\"ottingen Stellar Library (GSL, a theoretical library) and the first data release of MaStar, and external tests by comparing the resulting distributions in the parameter space with the APOGEE estimates of the same properties. In summary, our parameter estimates span in the ranges: $\teff=[2900,12000]\,$K, $\logg=[-0.5,5.6]$, $\FeH=[-3.74,0.81]$, $\alphaM=[-0.22,1.17]$. {We report internal (external) uncertainties of the properties of $\sigma_{\teff}\sim43\,(240)\,$K, $\sigma_{\logg}\sim0.2\,(0.4)$, $\sigma_{\FeH}\sim0.16\,(0.24)$, $\sigma_{\alphaM}\sim0.09\,(0.08)$.} These uncertainties are comparable to those of other methods with similar objectives. Despite the fact that \cosha{} is not aware of the spatial distribution of these physical properties in the Milky Way, we are able to recover the main trends known in the literature. The catalog of physical properties for MaStar can be accessed in \url{http://ifs.astroscu.unam.mx/MaStar}.

\end{abstract}



\section{Introduction}

The interpretation of galaxy observations into stellar physical properties ultimately relies on a set of ingredients, namely: a selection of stellar evolutionary tracks, an initial mass function and a stellar spectral library \citep[e.~g.,][]{Bruzual2003, Maraston2005, Conroy2009}. While, the first two are better constrained through our theoretical knowledge of the stellar interiors and stellar evolution, the latter provides the fundamental link between such theoretical knowledge and the observable quantities. In its essence, a stellar library is a collection of stellar spectra as homogeneous (in sampling and resolution) and curated (from instrumental and flux calibration artifacts) as possible, in the spectral space \citep[e.~g.,][]{Terndrup1990, Lancon1992, Fluks1994}. In the parameter space, usually defined by \teff, \logg, \FeH, and \alphaM, the distributions of stars in the library are expected to span the range of plausible physics according to the theory, also in an homogeneous way.

These requirements present several challenges since stellar surveys are mostly limited to our galaxy and, in some cases, further limited to the solar neighborhood \citep[e.~g.,][]{LeBorgne2003, Valdes2004, Sanchez-Blazquez2006, Chen2014}. Such limitations in our ability to acquire large samples of stars, eventually lead to a poor sampling of the parameter space. Furthermore, these limitations are exacerbated by the fact that most galaxies (including our own) usually display heterogeneous distributions in their stellar contents \citep[e.~g.,][]{Hayden2015, Fernandez-Alvar2017, Helmi2018, Barbuy2021}. {Therefore it is of paramount importance to better sample our own galaxy and eventually other galaxies with a large spatial coverage and a high spectral resolution to be able to measure spectral line abundances \citep[e.~g.,][]{GarciaPerez2016}.} To mitigate these restrictions in the parameter space sampled by {the so called \emph{empirical} stellar libraries compiled from observations}, several groups of authors have implemented \emph{theoretical} stellar libraries \citep[e.~g.,][]{Lejeune1997, Lejeune1998, Coelho2007, Meszaros2012}. Conceptually, theoretical libraries should be able to overcome the issue of sampling in parameter space. {However in practice, this flexibility comes at a cost: spectral model imperfections. Theoretical libraries can only model a limited observable space, mainly due to incomplete theoretical knowledge about stellar atmosphere opacities and/or incomplete atomic and molecular data \citep[see][and references therein]{Conroy2013a, Coelho2020}.}

Both empirical and theoretical libraries are clearly complementary approaches \citep[see][]{Coelho2020}. As a matter of fact, most commonly used synthesis of stellar population models combine both, in order to improve the accuracy in their predictions \citep[e.~g.,][]{CidFernandes2014}. In this sense, the MaNGA Stellar Library \citep[MaStar][]{Yan2019} comprising $24.4\,$k stars in the same footprint of the MaNGA survey, promises to be a huge leap forward in the direction of a better sampling of the parameter space while preserving the completeness of spectral features in the optical range of empirical libraries {if compared with other widely used libraries like MILES \citep{Sanchez-Blazquez2006}. Furthermore, MaStar poses the possibility of analyzing MaNGA IFS galaxies with a stellar library observed and reduced with the same instrumental settings. Such possibility would reduce the sources of uncertainties in the analysis of physical properties extracted from MaNGA galaxies.}

A major challenge is to calibrate such data set in \teff{}, \logg{}, \FeH{} and \alphaM{}, {since a high-quality data set with a similar coverage in spectral and parameter space would be needed in order to fit a predefined model or train a ML algorithm. We could use a theoretical model but those are still far from being consistent with one another as they usually rely on \emph{ad hoc} assumptions with huge impact on stellar atmospheres \citep[e.~g.,][]{Coelho2020, Lancon2021}  Hence we have to make a compromise \emph{a priori} with the limitation of the chosen theoretical library. Other alternatives could be to use observed and well calibrated stars (e.g., MILES, APOGEE), but those are usually restricted in spectral/parameter space and often rely themselves on theoretical models}.

{More recently, given the nature of the highly non-linearity of this problem, a departure from conventional fitting techniques has flourished in the literature \citep[e.~g.,][]{Singh2006, Ness2015, Sharma2020, Ting2019}. For instance, \citet{Ness2015} fits a 2-degree polynomial function in parameter space to a small ($\sim500$) set of training stars. They use this as a generative model to infer the physical properties (labels) of stars in the APOGEE survey \citep{Holtzman2018}. A limitation of \citet{Ness2015} approach is that they assume a functional form to model the spectrum using the known labels in the training set, also under the supposition that this labels are exhaustive in describing the stellar spectra and that the set functional model is accurate. To overcome those limitations, \citet{Ting2019} trained an artificial neural network (ANN), dubbed \textsc{The Payne} with a similar objective: to build a generative model and predict stellar parameters from observed spectra. Even though they managed to train with a relatively small sample of ($\sim2000$) stars, as a consequence their model is restricted in the parameter space. To mitigate such limitation, \citet{Xiang2021} introduced \textsc{HotPayne} to predict stellar abundances for observed hot stars ($>7500\,$K). Both, \citet{Xiang2015, Ting2019} are examples of machine learning approaches overcoming the limitation of imposing a functional form of the model \emph{a priori}, hence improving the likelihood of capturing non-linear behaviour when mapping from labels to spectral space. While training in the parameter space allowed the authors to substantially reduce the size demand on the training set when training on the spectral space, this advantage came at the expense of generalization of the model. Another limitation of both method (\textsc{cannon} and The Payne) is that both rely on spectral fitting to make the actual prediction of the labels, hence hampering the opportunity of using ML to its fullest potential, i.~e., by training in the spectral space to predict directly the labels. In this paper we seek to fill the gap of using conventional (non-ANN) ML approaches and to predict labels directly, without requiring spectral fitting.}

{Conventional machine learning (ML) algorithms (e.~g., ensemble methods) usually demand less training data than ANNs, while retaining the flexibility of the trained model not having a predefined shape \citep[e.~g.][see also Appendix~\ref{app:machine-learning}]{Geron2017}. We introduce a new method to \emph{directly} determine atmospheric physical parameters of the MaStar spectra by implementing a ML algorithm called Gradient Tree Boosting (GTB). GTB is an ensemble method that successively trains a predefined number of decision trees, each improving upon its predecessor errors. In early experiments we tested other ML regression methods, namely: decision tree, naive Bayes, support vector, etc.. Because our approach was heuristic from the beginning, we did not set to optimize hyper-parameters. We adopted GTB as was first method that returned reasonable results when tested on the testing set and compared to APOGEE.}

This paper is organized as follows. In \S~\ref{sec:train-test-sets} we describe the inputs for the training and testing sets; in \S~\ref{sec:cosha} we present the physical parameter estimator and training process; in \S~\ref{sec:model-evaluation} we evaluate the physical parameters using a set of internal and external tests, and in \S~\ref{sec:parameter-distributions} we present the MaStar stellar library physical properties; finally, we conclude in \S~\ref{sec:conclusions}.

\section{Training and testing sets}\label{sec:train-test-sets}

{In this section we describe the samples used to build the training and testing subsets.}

\begin{figure}
\centering
\includegraphics[width=\columnwidth]{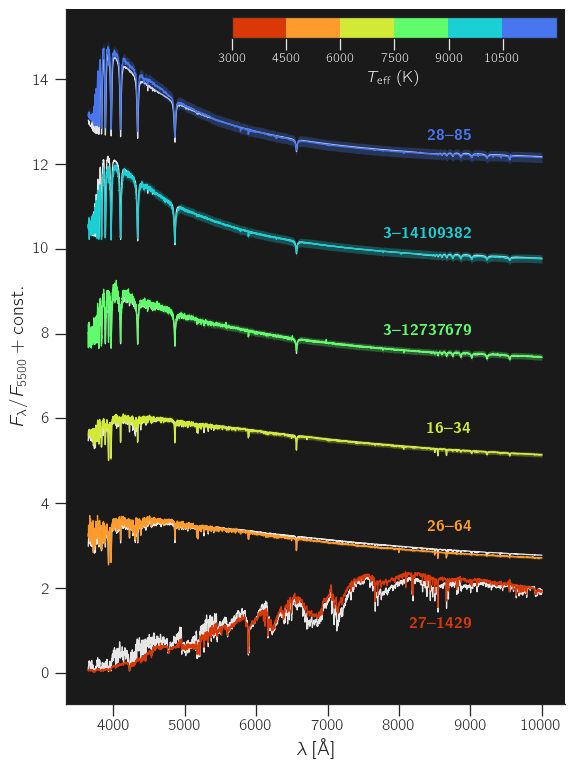}
\caption{{Examples of the spectra in the MaStar library color coded according to their \teff{} as predicted by \cosha{}. The typical $1\sigma$ band in each spectrum is also shown. In light grey we show the closest (in parameter space) GSL spectra to each MaStar spectrum. The mismatch in 27-1429 may be attributed to a mismatch in the actual properties of both spectra. The first data release of the library spans a wide range of the physical parameter space suitable for training a machine learning (ML) model to predict the properties of the rest of the stellar spectra.}}
\label{fig:spectra-showcase}
\end{figure}

\subsection{Input from the theoretical library GSL}\label{sec:theoretical-library}

Since we intend to use a ML method, we have to be aware that one of the most important challenges when it comes to this approach is the need for a clean, complete and already labelled sample \citep[e.~g.,][]{Geron2017}. In this particular case, in order to predict reliable estimates of \teff{}, \logg{}, \FeH{} and \alphaM{} for the MaStar spectra, we need a sample of stellar spectra with reliable estimates of these same properties. In the previous section we already sketch out the difficulties of having such empirical library at our disposal. However, state-of-the-art theoretical libraries do prove to be a suitable option.

In the literature there are plenty of theoretical libraries, most of which are focused on particular types of stars \citep[e.~g.,][]{Kirby2011, Coelho2014}. However, we require a multi-purpose theoretical stellar library in order to build models as general as possible. For this purpose we choose the G\"ottingen Spectral Library \citep[GSL;][$\sim27\,$k stars]{Husser2013}. The GSL is a stellar spectral library based on the latest version of the stellar atmospheric parameter code \phoenix{} \citep{Hauschildt1999}. This library offers the following advantages for our study: \textit{(i)} it samples a wide range of stellar stages, from sub-dwarf up to super giant stars ($\logg=0.0$~---~$6.0$), and spectral types ($\teff=2300$~---~$12000\,$K); \textit{(ii)} it predicts non-solar abundance patterns ($\FeH=-4.0$~---~$+1.0$ and $\alphaM=-0.2$~---~$+1.2$); \textit{(iii)} {it implements the latest atomic and molecular data, and opacity solutions to predict high-resolution spectra \citep[$R\sim500\,$k in $\lambda=3000$~---~$25000\,$\AA, see][and references therein]{Husser2013, Lancon2021}.} We acknowledge that perhaps two important weaknesses of the GSL are the lack of stars hotter than $12000\,$K and $\alpha$-enhanced/depleted atmospheres for stars outside the ranges $-3\leq\FeH\leq0\,$ and $3500\leq\teff\leq8000\,$K. For more details on the GSL, we refer the reader to \citet{Husser2013} and references therein.

In an attempt to mitigate some of the above-mentioned deficiencies of the GSL, we also adopt the physical properties included in the first data release of MaStar \citep[][$\sim3\,$k stars]{Yan2019}. The MaStar stellar properties presented in \citet[][hereafter MaStarDR1]{Yan2019} are a compilation of the properties reported by the APOGEE \citep{Majewski2017}, SEGUE \citep{Yanny2009} and LAMOST \citep{Cui2012} surveys, all of which are based on different kind of observations and analysis methods \citep[][respectively]{GarciaPerez2016, Lee2008a, Xiang2015}. These methods mostly implement algorithms based on template matching of spectroscopic and or photometric data to other (well-determined) observed or theoretical stellar libraries or a combination thereof. In particular, \citet{Lee2008a} also implements an Artificial Neural Network algorithm in part of their analysis to estimate atmospheric stellar properties. We stress that the atmospheric parameters reported in \citet{Yan2019} are intended to aid the target selection of the MaStar stellar library and not to be used on stellar population analysis. Furthermore, the fact that most of these stellar properties are derived from different wavelength ranges, signal-to-noise levels and using different types of analysis methods, we may expect different sources of uncertainties (not present in a theoretical library such as GSL) to propagate during the training stage.

\subsection{Input from the empirical library the MaNGA Stellar Library: MaStar}\label{sec:empirical-library}

{In this section we briefly describe describe the aspects of the MaStar library that are relevant to this paper.}

\subsubsection{Observations}\label{sec:observations}

MaNGA is an integral field spectroscopy survey of $10,000$ nearby galaxies \citep{Bundy2015, Law2015, Yan2016, Wake2017} as part of the SDSS-IV along with other two surveys: eBOSS \citep{Dawson2016} and APOGEE-2 \citep{Majewski2017}. It uses the Baryon Oscillation Spectroscopic Survey spectrographs \citep{Smee2013}, fiber-fed using different fiber bundle configurations for science targets, standard stars and sky observations \citep{Drory2015}. The wavelength sampling of the spectrographs combined covers $\sim2600$~---~$10000\,$\AA{} with an average resolution of $R=1800$. In order to make observations of different fields during one night, the fiber-plugged plates are stored in housings called cartridges. MaNGA shares six out of nine cartridges with APOGEE-2. This configuration presented the opportunity to piggyback on the APOGEE survey to build MaStar. The observing strategy, target selection and data reduction are detailed in \citet{Yan2019}.

\subsubsection{Quality control}\label{sec:quality}

\begin{figure*}
\centering
\includegraphics[width=0.9\textwidth]{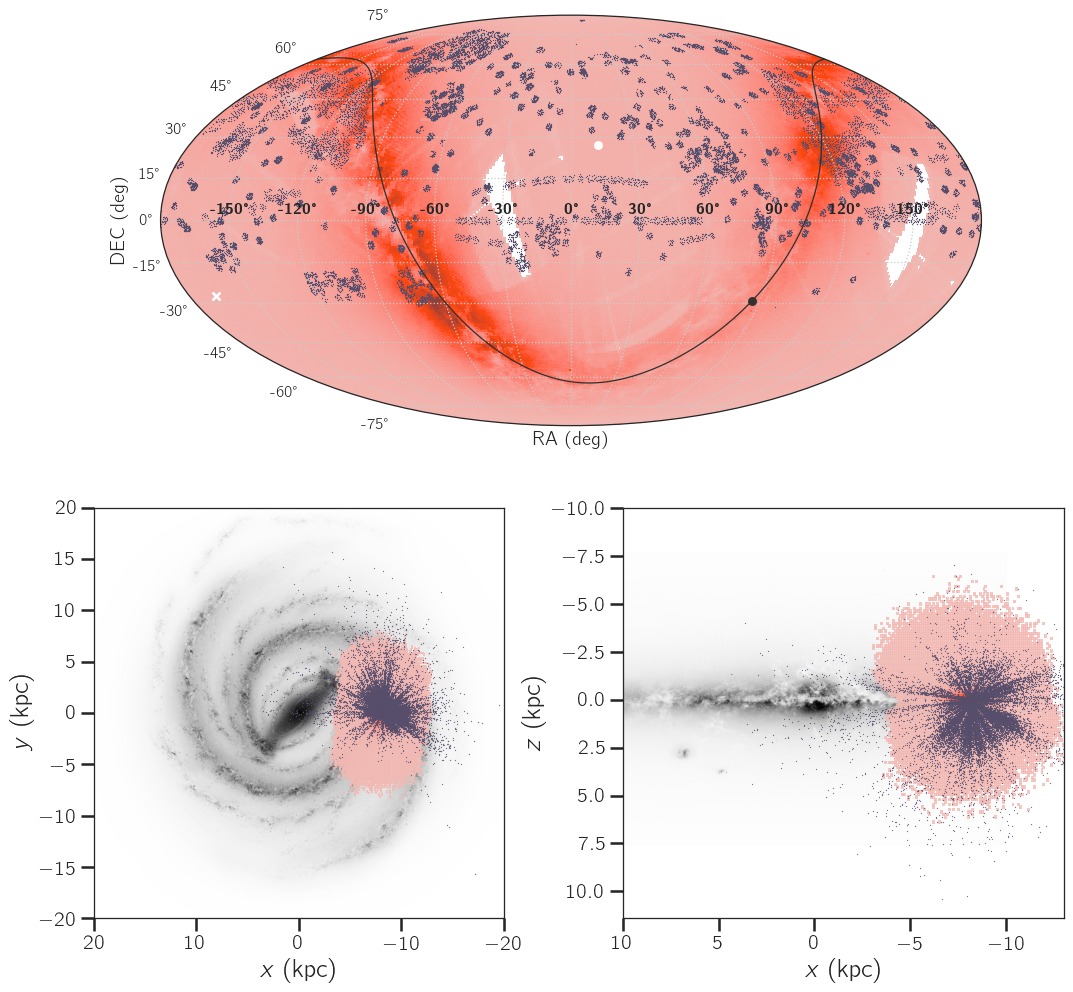}
\caption{The spatial distribution of the MaStar stars in the cleaned sample (purple) compared to that of the Gaia DR2 (orange). \textit{top:} the distribution in right ascension (RA) and declination (DEC) shows the in-homogeneous spatial sampling of the MaStar. \textit{bottom:} the distribution of MaStar in the galactic projections $x$~--~$y$ (left) and $x$~--~$z$ (right) demonstrates that MaStar extends well beyond the solar neighborhood. Lower panel plots were generated using the Python package \url{https://pypi.org/project/mw-plot/0.3.0/}.}
\label{fig:spatial-distribution}
\end{figure*}
{Before we can analyse the MaStar spectra, we need to ensure the quality of the observations. In this section we explain the cuts we take in order to have the best possible spectra to train the model.}

In the MaStar library, the stars have been observed under different atmospheric conditions, using different total integration times and most of them have been observed in several visits ($\sim3$ per star on average). In total there are $\sim68.2\,$k visits. It is expected that some visits to the same star may have better quality than others. In order to ensure that we keep only the best quality data, we use a series of internal quality flags and estimated parameters provided by \citetalias{Yan2019}, namely:

\begin{description}
\item[\texttt{MJDQUAL}] describes the quality of the spectroscopic calibration.
\item[\texttt{MNGTARG2}] describes the quality of the photometric selection.
\item[\texttt{RADVEL}] the radial velocity. Stars with no measure of this parameter are flagged as bad.
\end{description}

The first two are bit mask flags. \texttt{MJDQUAL} is required not to have bits $1$ (good sky subtraction), $4$ (good point spread function fitting), $5$ (good estimates of the extinction), $6$ (good estimation of the radial velocity, with scatter lower than $10\,$km~s$^{-1}$ across multiple exposures), $7$ (good calibration after visual inspection), $8$ (no emission line in H$\alpha$ with equivalent width greater than $0.6$\AA) and $9$ (good signal-to-noise per pixel, with median $>50$). \texttt{MNGTARG2} is required not to have bit $15$ activated, this means that the photometric selection is reliable. {To have good quality distances estimates from \citet{Bailer-Jones2018}, we further select only those stars in MaStar also present in Gaia DR2 \citep{Brown2018}. We implement these distance estimates to calculate dust extinction values $A_V$ for the MaStar spectra using the 3D dust map from \citet{GreenSchlafly2015} and the extinction curve model from \citet{Fitzpatrick1999} with $R_V=3.1$. The typical (median) extinction for MaStar is $A_V\sim0.1\,$mag.}

After selecting the best quality data according to the aforementioned flags, our sample comprises {$\sim23\,$k} good quality spectra of unique stars ($\sim98.5\%$ of the original sample). In Fig.~\ref{fig:spectra-showcase} we show several examples of the spectra in the MaStar library color-coded according to their \teff{} as reported in \citetalias{Yan2019}. {As a qualitative measure of how compatible MaStar and GSL spectra are, we also show the closest (in parameter space) GSL spectra in light grey. This comparison shows reasonable compatibility between both libraries. However, the observed differences, specially in the case of the coolest star, can be attributed to a mismatch in abundance parameters and/or the above mentioned limitations of theoretical libraries.}

{Even though we effectively clean the initial spectra sample from obvious artifacts using the internal flags described above, some spectra may still remain with notable artifacts due to imperfect flagging in the first place. There are several common issues still appreciated in some spectra: emission lines, noisy spectra, bad continuum calibration and missing pixels. Some of those problems could be addressed using external information, while others would remain being part of the limitations of the method.}

{In order to mitigate possible issues with the spectrophotometric calibration we use the Gaia DR2 \citep{Brown2018} colour distribution. We select only those stars from the MaStar library that have been catalogued by Gaia and that have a reliable astrometry and photometry \citep{Evans2018}. In addition, we derive the synthetic magnitude in the Gaia photometric system ($G_\text{bp}$, $G_\text{rp}$ and $G$) from the spectra of the MaStar matched library. Then we select those stars with: \textit{(i)} realistic blue Gaia color ($G_\text{bp}-G_\text{rp}\geq-0.76\,$mag), hence excluding stars with possible flux calibration issues; \textit{(ii)} a parallax $>0.05\,$mas to mitigate uncertain distances estimates; and \textit{(iii)} a median signal-to-noise ratio $S/N\geq20$ along the entire wavelength range. After cleaning the sample, we keep $\sim22\,$k stars with reliable flux calibration and peak near $S/N\sim100$, median $\sim140$ and mean $>200$ \citep[e.~g.,][]{Chen2020}. For the remaining of this study, we use this cleaned version of the MaStar sample, unless otherwise stated.}

In Fig.~\ref{fig:spatial-distribution} we show the spatial distribution of the MaStar stars, compared to that of Gaia. It is clear that MaStar presents a more incomplete coverage of the MW than Gaia, due to the nature of the former survey. None the less, it is clear that MaStar samples several regions of the Milky Way and is not limited to the solar neighborhood, within the declination limits of the survey.

\subsubsection{Pre-processing}\label{sec:preprocessing}

In this section we implement the ML algorithms. The unfamiliar reader may refer to \S~\ref{app:machine-learning} to briefly get acquainted on the common ML jargon and the notation adopted throughout this study. We encourage the seasoned reader to skip \S~\ref{app:machine-learning} altogether.

To solve the problems related to the presence of emission lines and the missing pixels, we implement two separate unsupervised ML algorithms: an outlier detection method and a missing features (pixels) filler. Both of these algorithms are based on the $k$-nearest neighbors algorithm \citep[KNN,][]{Fix1951}. In a nutshell, the KNN algorithm consists on finding the closer $k$ samples (e.~g., in an Euclidean sense) to each sample spectrum. 
In the following we elaborate on the usage of the KNN to remove emission lines and to fill-in missing pixels.

For the emission line problem, we implement the Local Outlier Factor (LOF) algorithm \citep{Breunig2000}, from the \textsc{scikit-learn} Python library \citep{Pedregosa2011}. The problem is posed as follows: given an observed spectrum $f_{\text{obs}}=(\lambda_i,f_i)$ for $i=1,\ldots,n$, this method looks for under-densities in the locality of each pixel $(\lambda_i,f_i)$. Such locality is defined by the $k$-nearest neighbors to $(\lambda_i,f_i)$ and the surrounding density is measured using the distance between those neighbors. In this case we define an Euclidean distance between each pair $(\lambda_i,f_i)$ and the rest of the spectrum pixels. A pixel is then flagged as outlier (and turned into a missing pixel) if its local density is smaller than a factor of the density of its $k$-neighbors. Given the nature of the stellar spectral energy distributions, this algorithm is prone to produce a large fraction of false positives (e.~g., absorption lines). In order to regularize the LOF algorithm, we set $k=5$ and the expected contamination for emission lines to be $0.4\%$. This way we improve the accuracy of the emission line detection. 

For the missing pixel problem, we implement the KNN ``imputer'' (filler) algorithm \citep{Troyanskaya2001}. Essentially, given a spectrum observed in the wavelength range $f_{\text{obs}}=(\lambda_i,f_i)$ for $i=1,\ldots,n$ with missing values in an arbitrary sub-subset of wavelengths $\lambda_{\text{missing}}\in\{\lambda_m\}\subset\lambda_{\text{obs}}$, this method looks for the KNN spectra in the library with no missing values in $\lambda_{\text{missing}}$ and then computes a distance-weighted average spectrum defined in $\lambda_{\text{missing}}$ to fill-in the missing values. 

\subsection{Training and testing sets}\label{sec:training-testing-sets}

Once we have cleaned the sample and pre-processed the spectra, we select the training and the testing subsets. {We arrange the spectra (\citetalias{Yan2019}+GSL) into a $N_{\text{spec}}\times N_{\text{pix}}$ matrix, so that each row is a spectrum and each column is a wavelength pixel, where $N_{\text{spec}}\sim30\,$k (\citetalias{Yan2019}: $2.7\,$k; GSL: $27\,$k) and $N_{\text{pix}}=6351$. In order to build such matrix, we resampled both libraries to $\Delta\lambda=1\,$\AA{} in the range $\lambda=3650~$--$~10000\,$\AA{} and downgraded the GSL spectral resolution to the \citetalias{Yan2019} wavelength resolution. The chosen wavelength sampling is arbitrary, but covers most of the original MaStar wavelength range. We further normalize the spectra to a common flux scale defined by $f_\lambda/f_{5500}$, so that only the shape of the spectra will inform the algorithm about the physical properties.} We will refer to this $N_{\text{spec}}\times N_{\text{pix}}$ matrix as the \emph{features matrix}, denoted by \fea{}. Correspondingly, each spectrum in the features matrix will be characterized by the set of parameters (\teff, \logg, \FeH, \alphaM), arranged into a $N_{\text{spec}}\times4$ \emph{labels matrix}, \lab{}. It is the goal of this study to train a model using the features matrix to translate new observations (rows in the features matrix) into its corresponding physical properties.

\begin{figure*}
\centering
\includegraphics[width=0.9\textwidth]{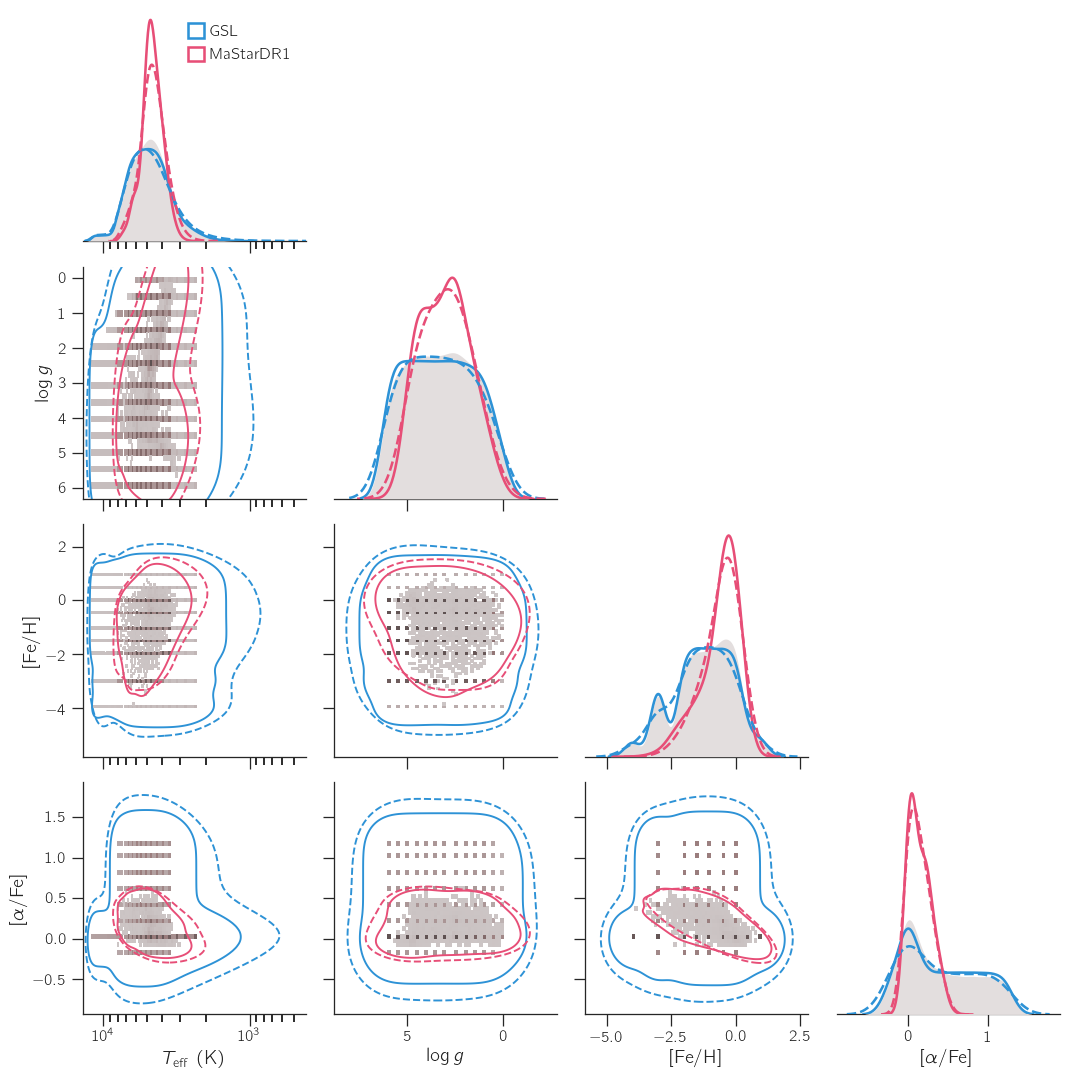}
\caption{We show the distribution of physical parameters (\teff{}, \logg{}, \FeH{} and \alphaM{}) in a set of paired diagrams and histograms for the full sample of $30\,$k star spectra with a good determination of the stellar properties (GSL+\citetalias{Yan2019}, grey values), together with the $99\%$ confidence region (contours) of the adopted training (solid) and testing (dashed) subsets corresponding to the GSL (blue) and \citetalias{Yan2019} (red) libraries. Clearly GSL spans a wider range of stellar properties than the MaStar DR1. See text in \S~\ref{sec:training-testing-sets} for details.}
\label{fig:training-testing-sets}
\end{figure*}
It is important that the training and the testing subsets are not biased against common types of stars (e.~g., dwarfs). This is to ensure that the trained model is as general as possible and can predict stellar properties of any kind of stars. From the stellar spectra libraries discussed in the \S~\ref{sec:train-test-sets}, only GSL fulfills these requirements, albeit with the intrinsic limitations of a synthetic stellar library. Therefore, we select a random subset comprising $90\%$ {($\sim25\,$k) of the stars in the GSL to be part of the training set. In order to avoid hampering the estimator during training with stellar spectral lacking the common instrumental and calibration artifacts (e.~g., low $S/N$, emission lines, sky subtraction and flux calibration), we also include as part of the training a random subset of $90\%$ {($\sim2.4\,$k)} of the MaStar DR1 spectra. The training subset comprises $\sim27\,$k out of the total of $\sim30\,$k stellar spectra with good estimation of the physical parameters. The remaining $10\%$ ($\sim3\,$k) of the GSL+\citetalias{Yan2019} ($\sim2.7+0.3\,$k)} stars are devoted to test the parameter estimator after training. Since most of the stars in these subsets have well known physical properties, the testing selection is suitable for the set purposes. In Fig.~\ref{fig:training-testing-sets} we show the distribution of the parameters in the GSL+\citetalias{Yan2019} sample (grey), and the corresponding segregation by stellar library (GSL: blue, \citetalias{Yan2019}: red) and by subset (training: solid line, testing: dashed line), both represented by $99\%$ confidence contours. As expected the GSL stellar library follows a nearly uniform distribution across the parameter space, while the \citetalias{Yan2019} draws a distribution that resembles the observed trends in stellar populations (e.~g. in the solar neighborhood). The fact that the training and testing subsets have similar distributions, regardless of the stellar library, is indicative that the testing procedure will be robust, despite the much smaller size of this subset.

In the next section we elaborate on the algorithm adopted for the parameter estimator, the training and the testing procedures. We will refer to the corresponding subsets as $\fea[\text{train}]$, $\lab[\text{train}]$ and $\fea[\text{test}]$, $\lab[\text{test}]$, respectively.



\section{The properties estimator: \cosha{}}\label{sec:cosha}

Most ML algorithms can be mathematically expressed as a functional $\mathcal{F}({\boldsymbol\theta},{\boldsymbol\phi})$ where $\boldsymbol\theta$ is a vector of parameters that define the trained (fitted) model and $\boldsymbol\phi$ is the vector of hyper-parameters that define how the model will be trained (e.~g., the merit or loss function, optimization algorithm, etc.). Once the hyper-parameters are set, $F\equiv\mathcal{F}({\boldsymbol\theta},{\boldsymbol\phi})$, the algorithm is ready to train a model $\hat{F}=F(\fea[\text{train}],\lab[\text{train}])$. The training process to build $\hat{F}$, consists in maximizing (minimizing) the score (loss) function in order to find the optimal set of parameters, $\hat{\boldsymbol\theta}$. Once trained, the fitted model should be able to provide reliable predictions given a set of features from new observations (not seen during the training). In reality, most problems require further exploration of the hyper-parameter space during the training phase in order to ensure robust results. However, in order to save computational time, we adopt an heuristic approach to train an \emph{ad-hoc} selection of hyper-parameters \citep[see e.~g.,][]{Geron2017, Ivezic2019}.

{In choosing our preferred ML algorithm, we distinguish between conventional ML (e.~g., decision trees, support vector machine) and Deep Learning (e.~g., ANNs) algorithms. Although the latter are becoming increasingly frequent in the astronomical community \citep[e.~g.,][]{Ting2019, Xiang2015}. One important caveat of ANNs is that they usually require high computational power and/or large training samples to reach a desired level of accuracy without compromising generalisation \citep[e.~g., ][]{Chollet2017}. There are ways to overcome these limitations when training an ANN, namely: reducing the number of neurons/layers in the ANN, change the architecture of the ANN (e.~g., by limiting the number of connections per neurons) and/or constraining model predictions to a smaller label/feature space. Even though these ways may effectively reduce training time required to reach a robust ANN prediction, they impose a trade-off with the generalization of the model \citep[e.~g.,][]{Ting2019}.}

{In this study we implement a conventional ML algorithm, which require less training data and computational power, whilst retaining model flexibility and generalization \citep[e.~g.,][]{Geron2017}. We therefore choose a Gradient Tree Boosting (GTB) algorithm to train a model suitable for predicting the physical properties of the MaStar spectra. The GTB has two interesting characteristics, namely: it is an ensemble method and it is based on decision trees. Given the relatively small training set ($27.4\,$k spectra), a decision tree based method is a convenient option. We choose to implement our code based in the package \textsc{scikit-learn} \citep{scikit-learn}. This allow to keep our code base small, as we only have to implement the data munging and book-keeping procedures, namely: reading the sample, pre-processing, splitting into training and testing, storing the trained model and storing its predictions. Another important advantage of the \textsc{scikit-learn} of GTB implementation is that it allows for quantile regression, a type of regression in which it is possible to estimate any quantile of the probability distribution of \lab[] conditional on \fea[], $P(\lab[]\,|\,\fea[])$. Hence, this type of regression is more general as it is not limited to finding the mean of that distribution, $\hat{F}$. In order to train a quantile regression estimator, we change the hyper-parameters of the GTB algorithm accordingly and set to predict the $16$th and $84$th percentiles (P$_{16}$ and P$_{84}$, respectively). In Table~\ref{tab:hyperparameters} we show the hyper-parameters used to train the different estimators. In the following we use the \emph{mode} estimator to predict the stellar labels, unless otherwise stated.}

\begin{table}
\centering
\caption{Hyper-parameters for \cosha{}.}
\label{tab:hyperparameters}
\begin{tabular}{lcccc}
\hline
& mode & P$_{16}$ & P$_{50}$ & P$_{84}$ \\
\hline
\verb"learning_rate" & 0.1 & 0.1 & 0.1 &  0.1  \\
\verb"loss" & ls & quantile & quantile & quantile  \\
\verb"alpha" & - & 0.16 &  0.50 &  0.84   \\
\verb"max_depth" & 4 & 4 &  4 &  4   \\
\verb"min_samples_split" & 5 & 5 & 5 &  5 \\
\verb"n_estimators" & 100  & 100 &  100 & 100 \\
\hline
\end{tabular}
\end{table}

Moreover, the fact that GTB is an ensemble algorithm means that it is the combination of several individual estimators (in this instance, decision trees). The boosting character of this algorithm comes from the fact that each trained decision tree improves its predecessor. This last trait of the GTB algorithm entails more robust results than single-estimator algorithms. A GTB is therefore a type of ensemble composed by a predefined number of $n$ decision trees, that can be expressed as the summation:
\begin{equation}\label{eq:gtb-estimator}
\hat{F} = f_0(\fea[\text{train}],\lab[\text{train}]) + \sum_{i=1}^{n-1}f_{i}(\fea[\text{train}],\Delta\lab[i]),
\end{equation}
where $f_0$ is the first decision tree trained on the original training subset (\fea[\text{train}] and \lab[\text{train}]) and the successive $f_i$ ($i>0$) are the decision trees trained on the original training spectra (\fea[\text{train}]) but using as labels the residual between the original labels and the label predicted by the previous estimator, $\Delta\lab[i]\equiv\lab[\text{train}]-\hat{f}_{i-1}(\fea[\text{train}])$.

We acknowledge GTB carry some caveats, {for instance, since it is built upon the combination of several trained models, the easiness in the interpretation of the model is mitigated, if compared to a decision tree. Still GTB methods are more easy to interpret than ANN \citep[e.~g.,][]{Chollet2017}. Furthermore, ANNs' flexibility implies that a large hyper-parameter space needs to be defined/constrained: number of layers, number of neurons per layer, number of neuron connections, type of connection response, etc.. From Table~\ref{tab:hyperparameters}, the hyper-parameters that we actually change during the development of this work are the \Verb"loss" and \Verb"alpha". The rest of the parameters were left to their default values according to \textsc{scikit-learn} \Verb"v0.23.2".} 

Finally, we use the trained model in $\hat{F}$ to predict the stellar properties from the entire clean spectral library, to build the matrix \lab[\text{model}]. We also use the estimators trained for the percentiles P$_{16}$ and P$_{84}$ to predict the precision of \cosha{} by computing $\sigma_{\lab[]}=(\text{P}_{84}-\text{P}_{16})/2$, which under the assumption of a $P(\lab[]\,|\,\fea[])$ Gaussian distribution is equivalent to the standard deviation. In the following sections we evaluate the internal reliability of the resulting distributions in \lab[\text{model}] in the stellar parameter space.

\section{Model evaluation}\label{sec:model-evaluation}

In this section we evaluate the reliability of the method described in the previous sections. We run two types of tests: internal and external.\footnote{{In summary, internal uncertainty refers to those uncertainties we draw from comparing with the testing sample, which is consistent with the samples used during training. External uncertainty (or consistency rather) are those we draw from comparing with external catalogs: APOGEE-ASPCAP/CANNON, MILES, etc. We expect that internal uncertainty to be smaller than external, since different sources of uncertainties combine when comparing with other catalogs: methods, observations, etc.}} In the internal tests we look for the model accuracy on the testing set. We also look for trends between the residuals of our method and the true values in this subset.

Quantitatively, we measure the accuracy of the model through the residual defined as:
\begin{equation}\label{eq:residual}
\Delta \lab{} \equiv \lab[\text{model}] - \lab[\text{true}],
\end{equation}
where $\lab[\text{model}]\equiv\hat{F}(\fea)$, \fea{} is any set of spectral features for which we know the corresponding set of true properties, $\lab[\text{true}]$. We adopt the mean and the standard deviation of these residuals as an estimation of the systematic and random errors, respectively.


In the external tests, on the other hand, we compare the predictions of \cosha{} with {the previously published results}, namely: {the MaStar DR1 (\citetalias{Yan2019}) and APOGEE DR14 \citep{Majewski2017, Holtzman2018}.} We restricted this comparison to those stars covered by both samples. This comparison only accounts for the consistency between the methods being compared, and not the absolute accuracy of the procedure. We estimate the level of consistency ($\delta_{\text{other}}$) through the definition of the discrepancy (not to be confused with the residual above) between the two predictions, as:
\begin{equation}\label{eq:discrepancy}
\delta_{\text{other}}\,\lab{} \equiv \lab[\text{model}] - \lab[\text{other}],
\end{equation}
where $\lab[\text{other}]$ is the estimation performed by other authors on the same sample of stars.

Since the \citetalias{Yan2019} data set is part of the training/testing subsets, we reckon this is not an independent nor fair comparison. Nonetheless, the consistency between \cosha{} predictions and those of \citetalias{Yan2019} is still interesting and deserves some exploration.

\subsection{Accuracy}\label{sec:accuracy}

\figsetstart
\figsetnum{4}
\figsettitle{Precision and accuracy versus $S/N$}

\figsetgrpstart
\figsetgrpnum{4.1}
\figsetgrptitle{Joint distribution of residuals}
\figsetplot{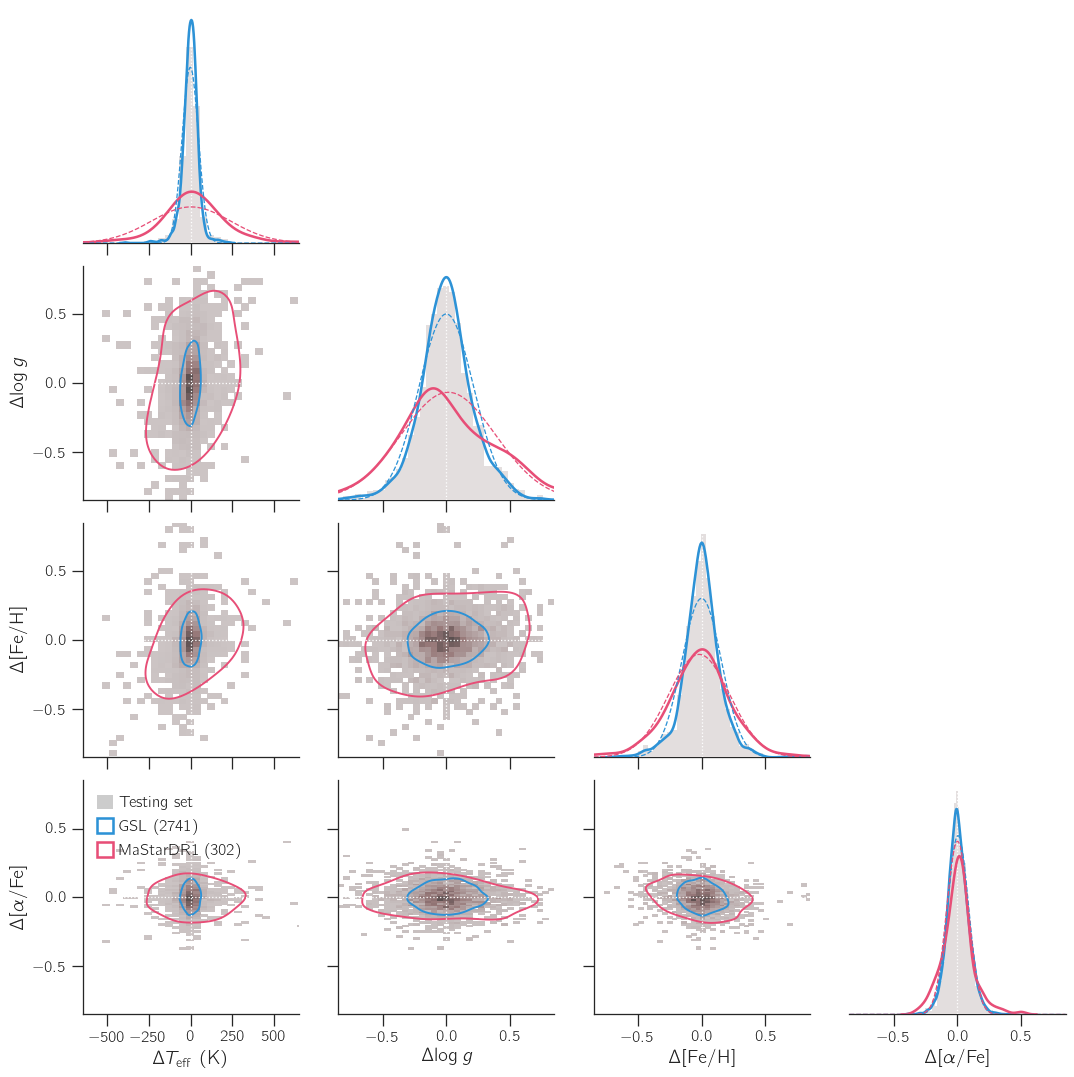}
\figsetgrpnote{Joint distribution of residuals as defined in Eq.~\eqref{eq:residual} in the parameter space. The results for the testing set is shown in grey. The residuals for the GSL and \citetalias{Yan2019} subsets are shown in blue and red, respectively, like in Fig.~\ref{fig:training-testing-sets}. The univariate residuals are represented in the diagonal planes (histograms and solid lines). A Gaussian distribution with the intrinsic mean and standard deviation of the residuals is also represented (dashed lines) in the diagonal planes. The contours in the off-diagonal planes enclose $1\sigma$ of the probability distribution.}
\figsetgrpend

\begin{figure*}
\centering
\includegraphics[width=0.9\textwidth]{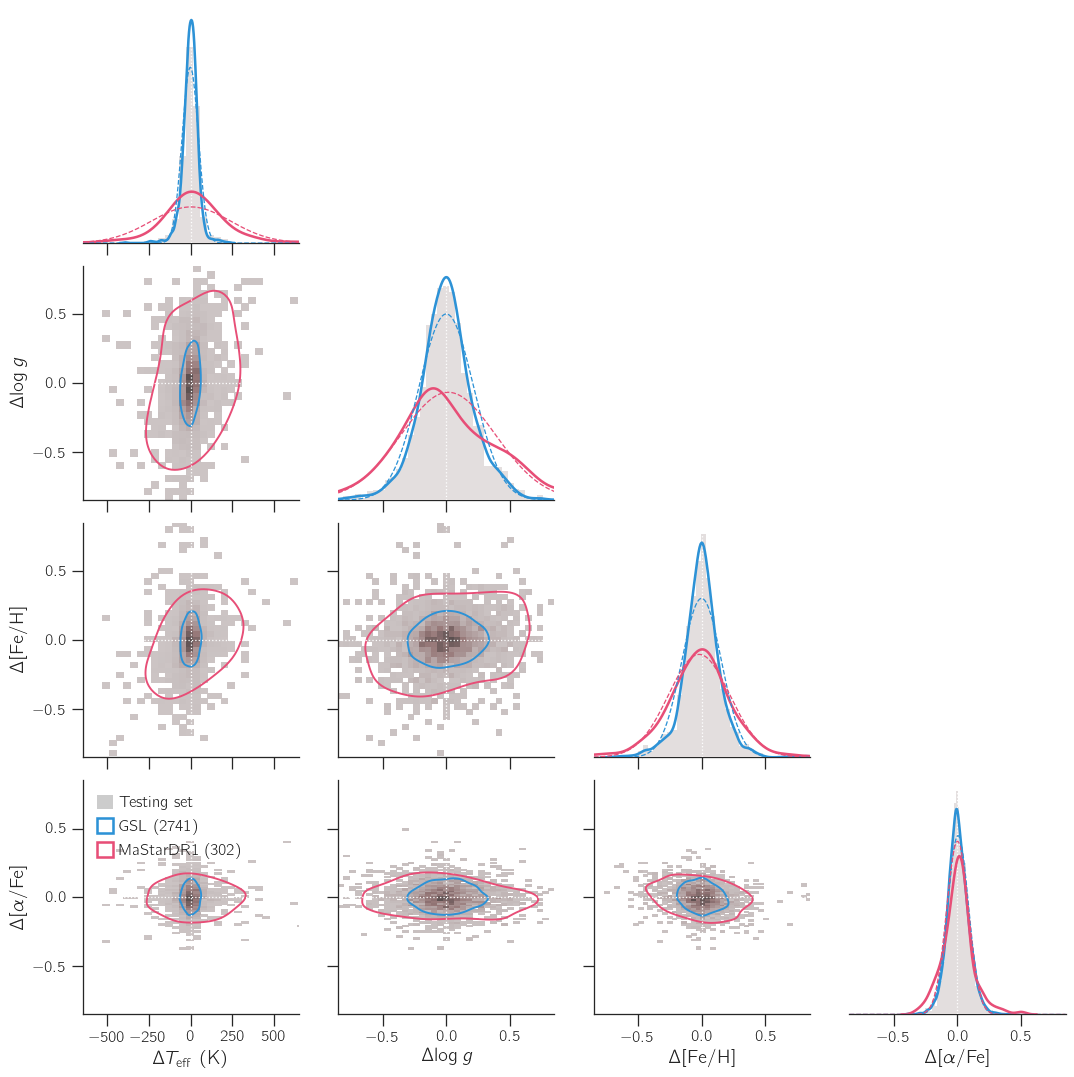}
\caption{Joint distribution of residuals as defined in Eq.~\eqref{eq:residual} in the parameter space. The results for the testing set is shown in grey. The residuals for the GSL and \citetalias{Yan2019} subsets are shown in blue and red, respectively, like in Fig.~\ref{fig:training-testing-sets}. The univariate residuals are represented in the diagonal planes (histograms and solid lines). A Gaussian distribution with the intrinsic mean and standard deviation of the residuals is also represented (dashed lines) in the diagonal planes. The contours in the off-diagonal planes enclose $1\sigma$ of the probability distribution. We note the intrinsic errors from \citetalias{Yan2019} are larger than our \cosha{} internal errors and show larger deviation from the Gaussian behaviour. The lack of correlation in almost all projections of the residuals space suggests a striking lack of degeneracies. See text in \S~\ref{sec:accuracy} for details. The complete figure set (9 images) is available in the online journal and includes similar figures of the precision and accuracy versus S\/N.
}
\label{fig:consistency-testing}
\end{figure*}

\figsetgrpstart
\figsetgrpnum{4.2}
\figsetgrptitle{Joint distribution of residuals for giant stars}
\figsetplot{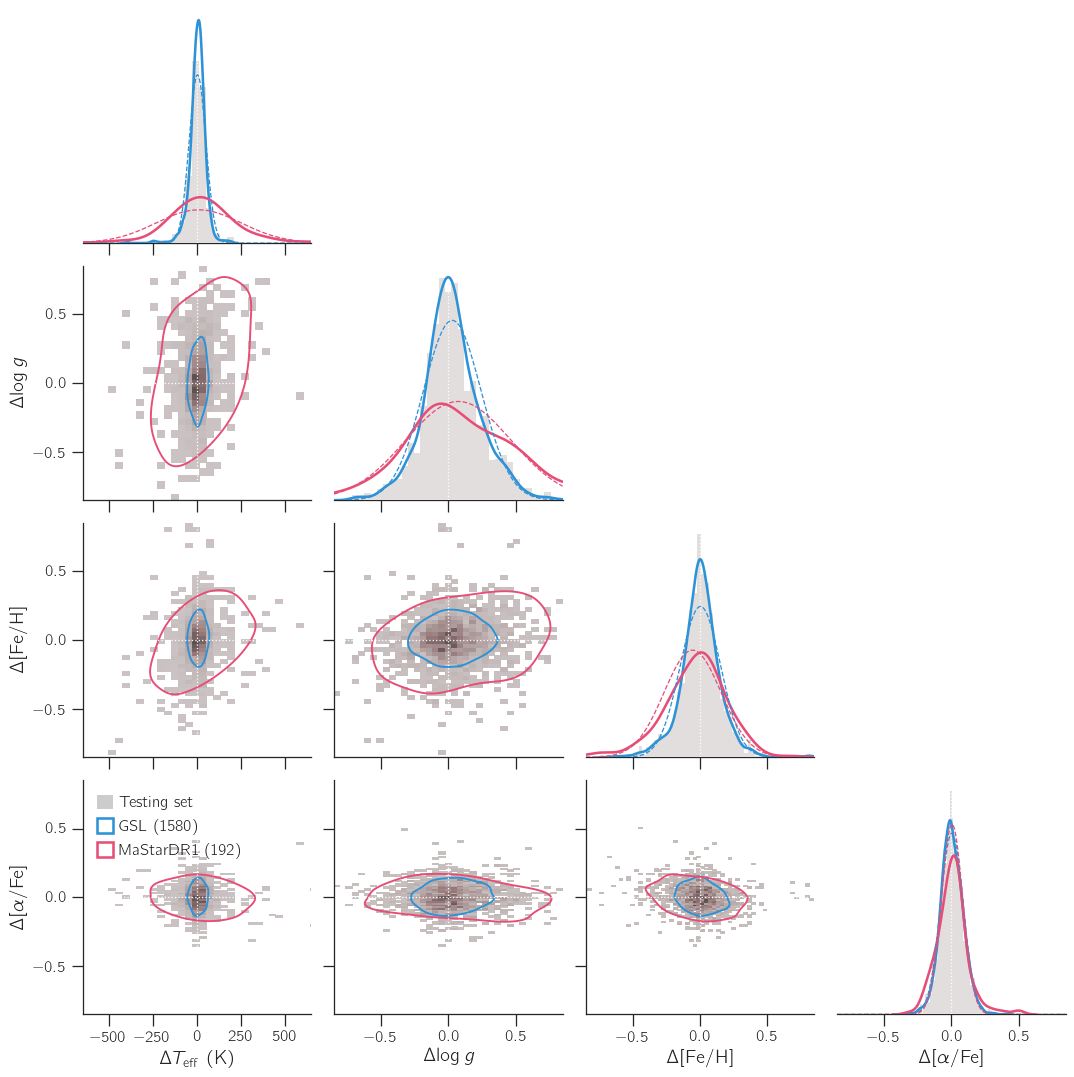}
\figsetgrpnote{Joint distribution of residuals for giant ($\logg\leq3.5$) stars in the testing sample. The results for the testing set is shown in grey. The residuals for the GSL and \citetalias{Yan2019} subsets are shown in blue and red, respectively, like in Fig.~\ref{fig:training-testing-sets}. The univariate residuals are represented in the diagonal planes (histograms and solid lines). A Gaussian distribution with the intrinsic mean and standard deviation of the residuals is also represented (dashed lines) in the diagonal planes. The contours in the off-diagonal planes enclose $1\sigma$ of the probability distribution. As in Fig.~\ref{fig:consistency-testing}, we find no correlation in GSL,  \citetalias{Yan2019} subsets.}
\figsetgrpend

\figsetgrpstart
\figsetgrpnum{4.3}
\figsetgrptitle{Joint distribution of residuals for dwarf stars}
\figsetplot{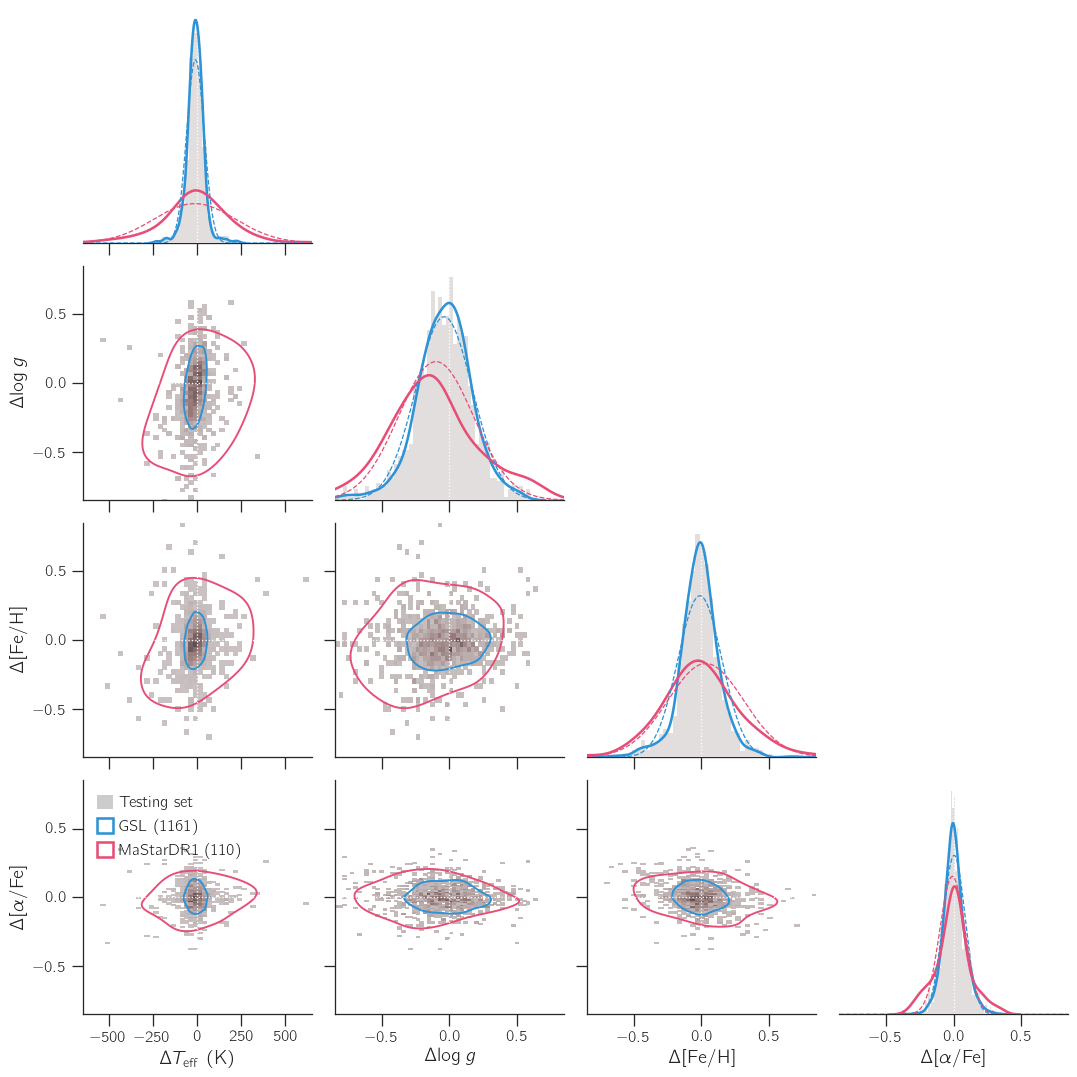}
\figsetgrpnote{Joint distribution of residuals for dwarf ($\logg>3.5$) stars in the testing sample. The results for the testing set is shown in grey. The residuals for the GSL and \citetalias{Yan2019} subsets are shown in blue and red, respectively, like in Fig.~\ref{fig:training-testing-sets}. The univariate residuals are represented in the diagonal planes (histograms and solid lines). A Gaussian distribution with the intrinsic mean and standard deviation of the residuals is also represented (dashed lines) in the diagonal planes. The contours in the off-diagonal planes enclose $1\sigma$ of the probability distribution. We find no correlation in GSL as in Fig.~\ref{fig:consistency-testing} and the existing correlations in \citetalias{Yan2019} subsets where apparently mitigated.}
\figsetgrpend

\figsetgrpstart
\figsetgrpnum{4.4}
\figsetgrptitle{Accuracy for $S/N=\infty$}
\figsetplot{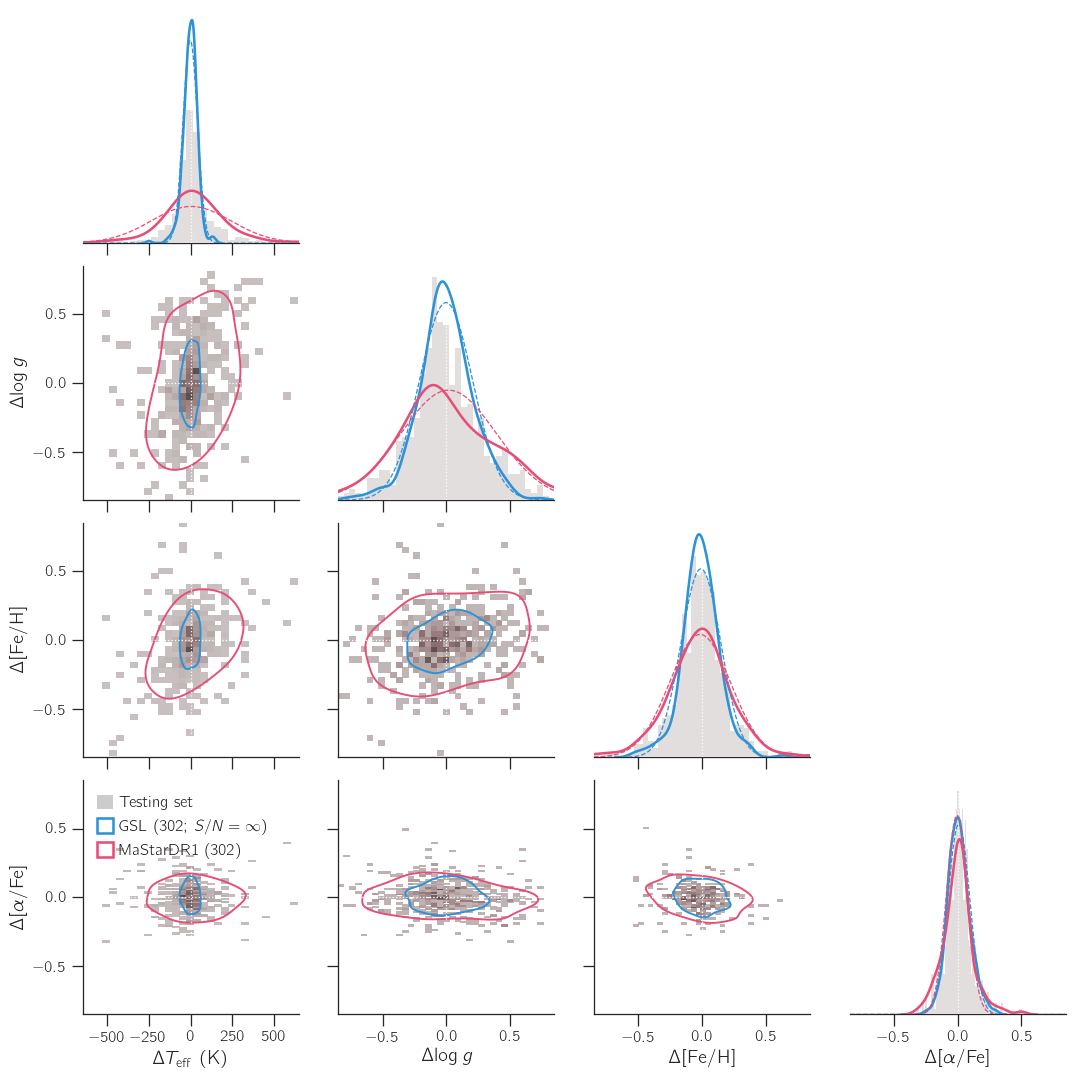}
\figsetgrpnote{Accuracy for $S/N=\infty$. The results for the testing set is shown in grey. The residuals for the GSL and \citetalias{Yan2019} subsets are shown in blue and red, respectively, like in Fig.~\ref{fig:training-testing-sets}. The univariate residuals are represented in the diagonal planes (histograms and solid lines). A Gaussian distribution with the intrinsic mean and standard deviation of the residuals is also represented (dashed lines) in the diagonal planes. The contours in the off-diagonal planes enclose $1\sigma$ of the probability distribution.}
\figsetgrpend

\figsetgrpstart
\figsetgrpnum{4.5}
\figsetgrptitle{Accuracy for $S/N=300$}
\figsetplot{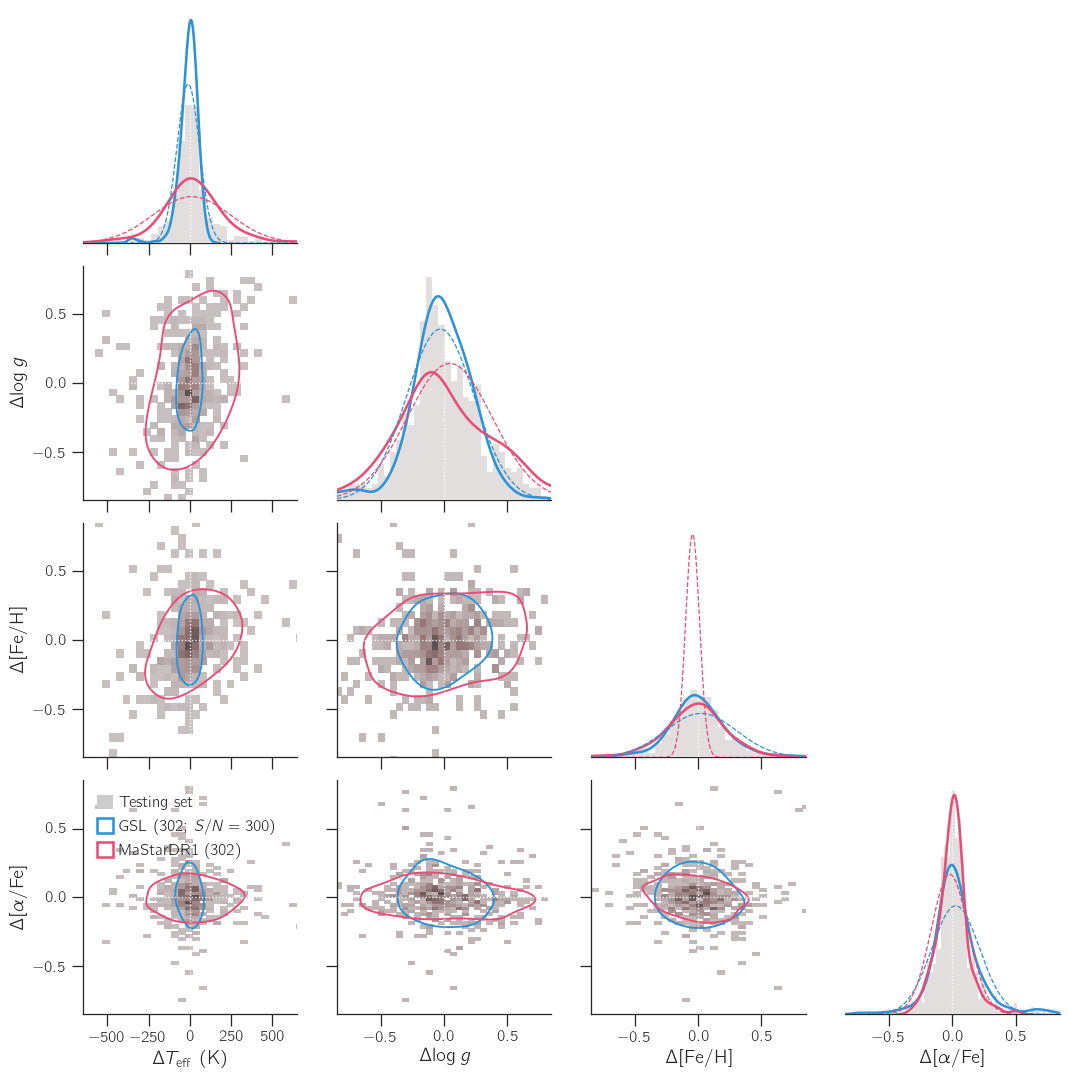}
\figsetgrpnote{Accuracy for $S/N=300$. The residuals for the GSL and \citetalias{Yan2019} subsets are shown in blue and red, respectively, like in Fig.~\ref{fig:training-testing-sets}. The univariate residuals are represented in the diagonal planes (histograms and solid lines). A Gaussian distribution with the intrinsic mean and standard deviation of the residuals is also represented (dashed lines) in the diagonal planes. The contours in the off-diagonal planes enclose $1\sigma$ of the probability distribution.}
\figsetgrpend

\figsetgrpstart
\figsetgrpnum{4.6}
\figsetgrptitle{Accuracy for $S/N=200$}
\figsetplot{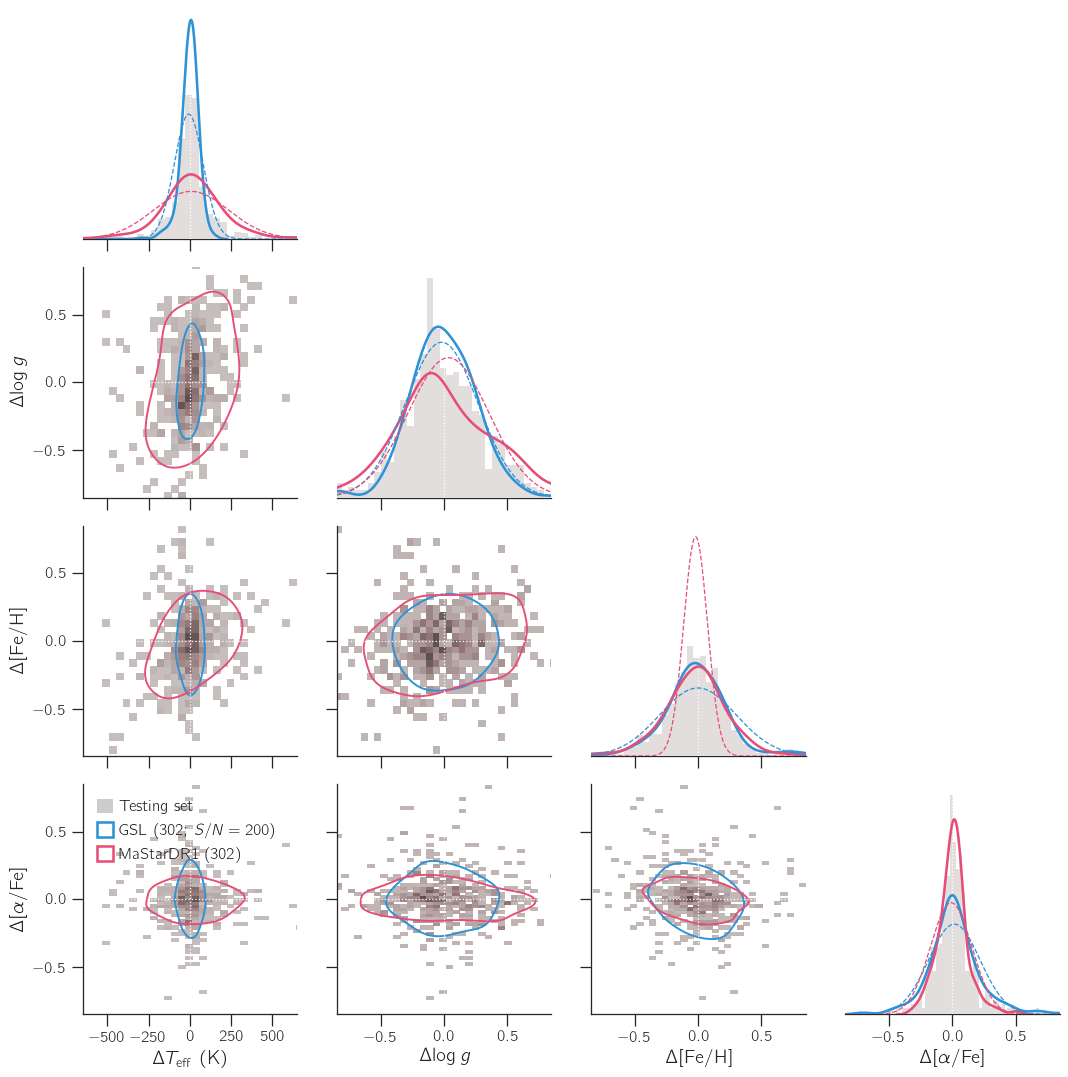}
\figsetgrpnote{Accuracy for $S/N=200$. The residuals for the GSL and \citetalias{Yan2019} subsets are shown in blue and red, respectively, like in Fig.~\ref{fig:training-testing-sets}. The univariate residuals are represented in the diagonal planes (histograms and solid lines). A Gaussian distribution with the intrinsic mean and standard deviation of the residuals is also represented (dashed lines) in the diagonal planes. The contours in the off-diagonal planes enclose $1\sigma$ of the probability distribution.}
\figsetgrpend

\figsetgrpstart
\figsetgrpnum{4.7}
\figsetgrptitle{Accuracy for $S/N=150$}
\figsetplot{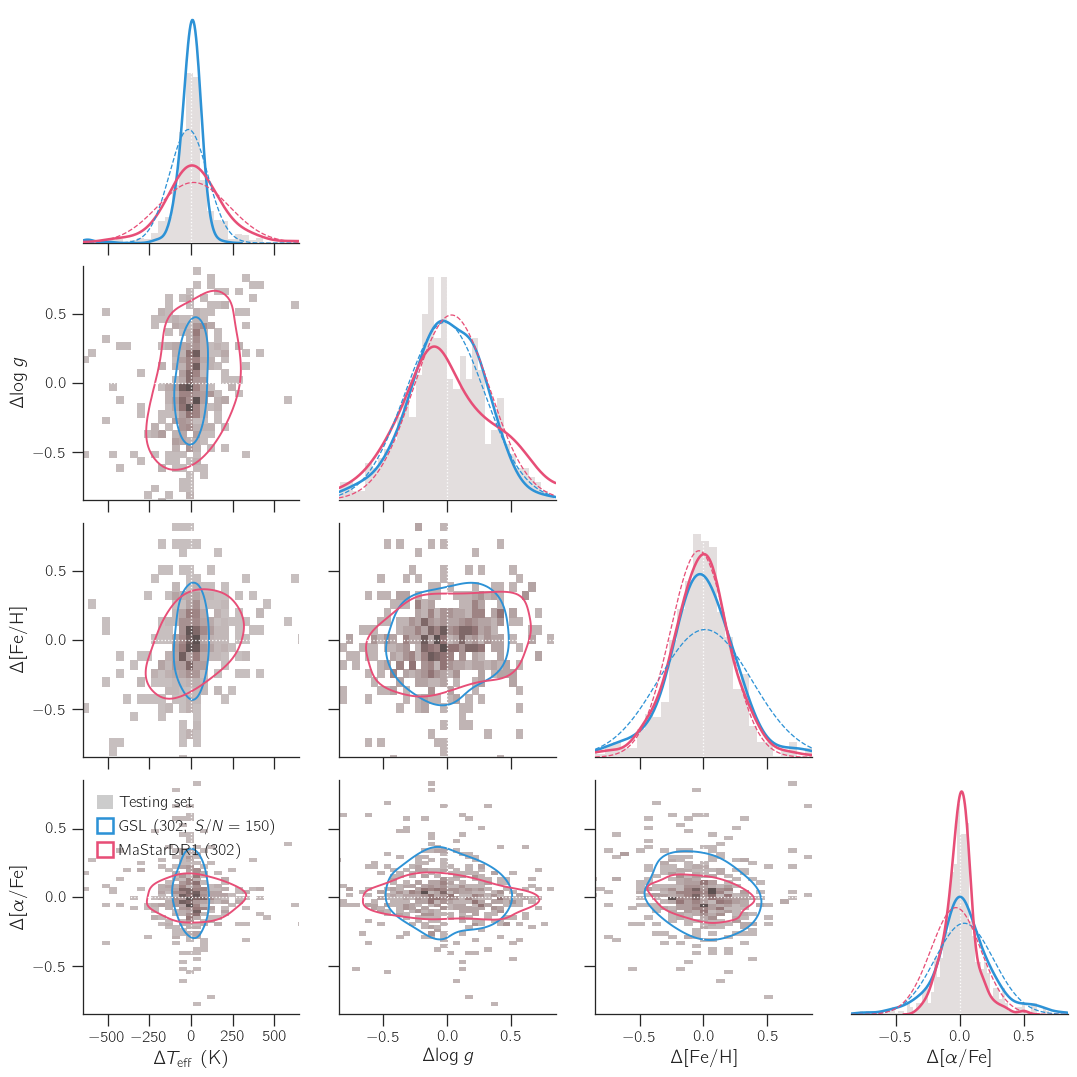}
\figsetgrpnote{Accuracy for $S/N=150$. The residuals for the GSL and \citetalias{Yan2019} subsets are shown in blue and red, respectively, like in Fig.~\ref{fig:training-testing-sets}. The univariate residuals are represented in the diagonal planes (histograms and solid lines). A Gaussian distribution with the intrinsic mean and standard deviation of the residuals is also represented (dashed lines) in the diagonal planes. The contours in the off-diagonal planes enclose $1\sigma$ of the probability distribution.}
\figsetgrpend

\figsetgrpstart
\figsetgrpnum{4.8}
\figsetgrptitle{Accuracy for $S/N=100$}
\figsetplot{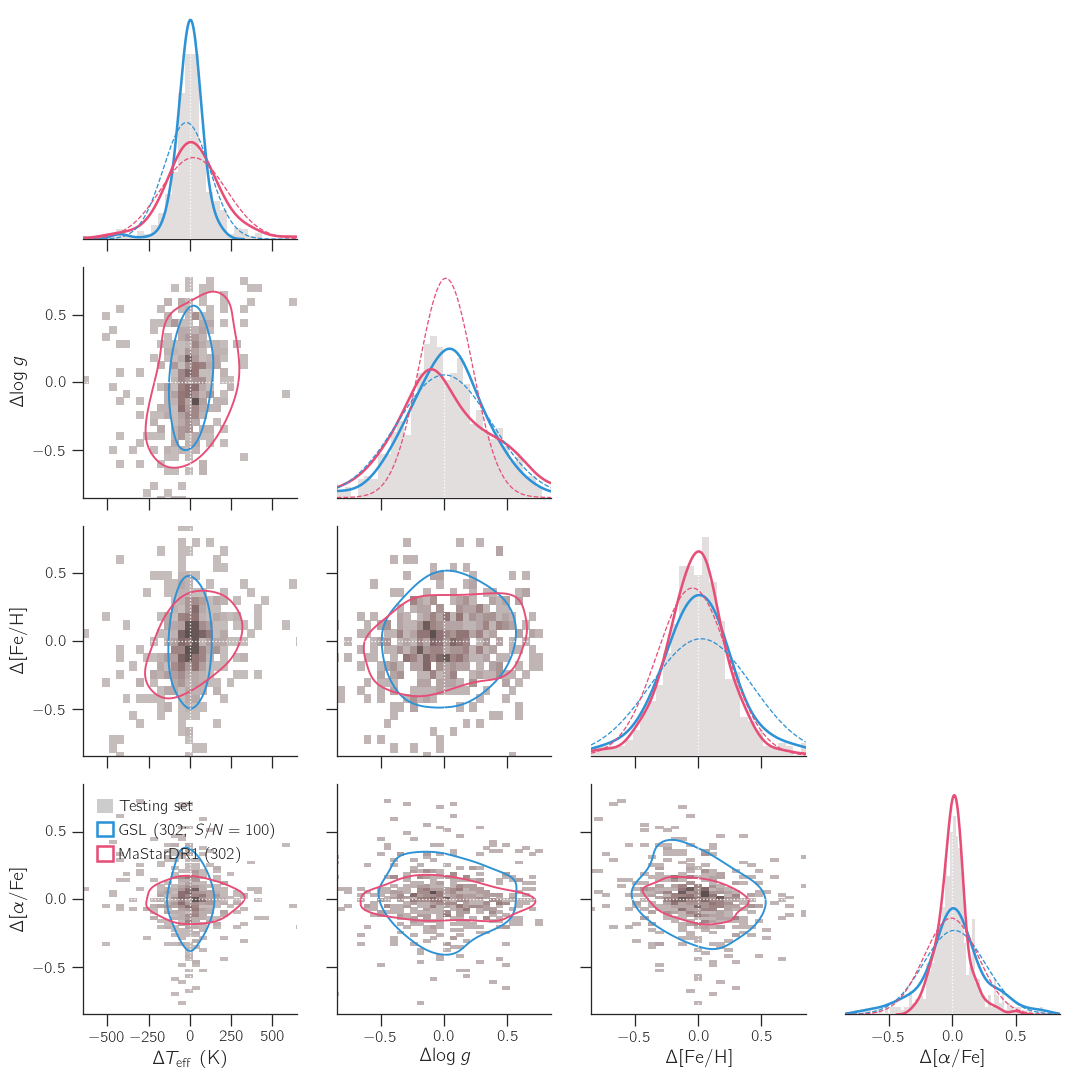}
\figsetgrpnote{Accuracy for $S/N=100$. The residuals for the GSL and \citetalias{Yan2019} subsets are shown in blue and red, respectively, like in Fig.~\ref{fig:training-testing-sets}. The univariate residuals are represented in the diagonal planes (histograms and solid lines). A Gaussian distribution with the intrinsic mean and standard deviation of the residuals is also represented (dashed lines) in the diagonal planes. The contours in the off-diagonal planes enclose $1\sigma$ of the probability distribution.}
\figsetgrpend

\figsetgrpstart
\figsetgrpnum{4.9}
\figsetgrptitle{Accuracy for $S/N=50$}
\figsetplot{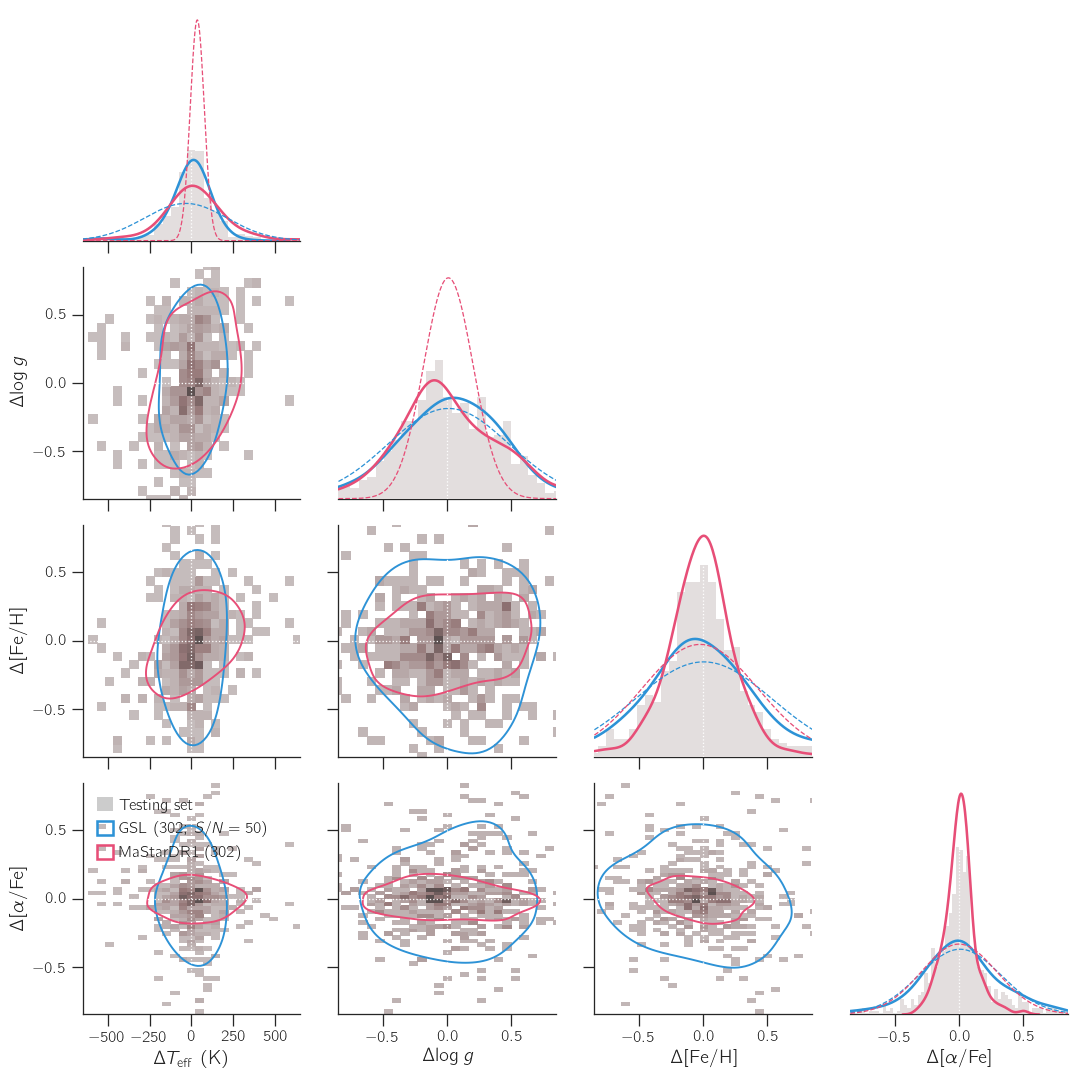}
\figsetgrpnote{Accuracy for $S/N=50$. The residuals for the GSL and \citetalias{Yan2019} subsets are shown in blue and red, respectively, like in Fig.~\ref{fig:training-testing-sets}. The univariate residuals are represented in the diagonal planes (histograms and solid lines). A Gaussian distribution with the intrinsic mean and standard deviation of the residuals is also represented (dashed lines) in the diagonal planes. The contours in the off-diagonal planes enclose $1\sigma$ of the probability distribution.}
\figsetgrpend

\figsetend

{In this section we use the testing subset of stars to measure the reliability of \cosha{}. Since a large percentage ($\sim90\%$) of this testing subset corresponds to theoretical predictions in the GSL, for these stars we can safely assume that we know the true values of the atmospheric properties studied in this work. Hence, we use the testing subset of GSL stars to quantify the \emph{internal} accuracy and precision of our method.} The rest of the testing subset corresponds to \citetalias{Yan2019}'s estimates and will be used to quantify the consistency between both methods. In Fig.~\ref{fig:consistency-testing} we show the residuals (c.~f. Eq.~\ref{eq:residual}) in the parameter space. The grey distributions represent the testing subset, while the GSL and \citetalias{Yan2019} subsets are represented (as in Fig.~\ref{fig:training-testing-sets}) in blue and red, respectively. The contours enclose $1\sigma$ of the corresponding probability distributions. We reckon that this kind of plot can reveal potential dependencies amongst the physical properties, or the lack thereof. These correlations are commonly known as degeneracies.

In general, all projections in the entire residual space (grey distributions) shows almost no correlations. {The accuracy is $\sim-1.4\,$K in \teff{} and $\sim0.004\,$dex (at most) for the rest of the properties.} However the segregation of the testing subset into GSL (solid blue) and \citetalias{Yan2019} (solid red) subsets uncovers that their individual accuracies are rather different. The comparison between these distributions (blue and red) shows that our method is both more accurate and precise than \citetalias{Yan2019}, across the parameter space. We recall though, that GSL residuals represent the real (internal) errors of our method while the residuals in the \citetalias{Yan2019} subset combines both \cosha{} and \citetalias{Yan2019} errors. We can estimate the \citetalias{Yan2019} intrinsic residuals as a Gaussian distribution with $\mu=\mu_\text{Y19} - \mu_\text{GSL}$ and $\sigma=\sqrt{\sigma_\text{Y19}^2-\sigma_\text{GSL}^2}$ (dashed red). Since the Gaussian supposition has the potential to visually hint asymmetries and other non-Gaussian behaviors, we also show, for the sake of completeness, the corresponding Gaussian distribution for the GSL subset (dashed blue), using $\mu=\mu_\text{GSL}$ and $\sigma=\sigma_\text{Y19}$. The residuals with the GSL shows the accuracy and precision across the parameter space characterized by $\Delta\teff\sim -3\pm48\,$K, $\Delta\logg\sim 0.00\pm0.20\,$, $\Delta\FeH\sim 0.00\pm0.13\,$ and $\Delta\alphaM\sim 0.00\pm0.09\,$. On the other hand for \citetalias{Yan2019}, the \emph{intrinsic} {accuracy and precision are, in general, larger: $\Delta\teff\sim 3\pm240\,$K, $\Delta\logg\sim 0.02\pm0.38\,$, $\Delta\FeH\sim -0.02\pm0.24\,$ and $\Delta\alphaM\sim -0.01\pm0.08\,$. Since these residuals and, in particular those quoted for the GSL testing set, are small we can safely rule out \cosha{} is over-fitting the training data. Otherwise, these residual distribution would display large biases.}

{In order to investigate how the accuracy behaves as a function of the noise level in the spectrum, we run the following experiment. In the testing subset of GSL spectra ($\sim2.7\,$k), we add random Gaussian noise at several $S/N$ levels ($50$, $100$, $200$, $300$ and $\infty$) and predict in each case the corresponding stellar parameters using \cosha{}. In Table~\ref{tab:prec-acc} we show a summary of our results. We find that the accuracy is independent of the level of noise of the input spectra. This further supports our conclusion in the last paragraph: \cosha{} is not suffering from over-fitting. We complement these results with figures similar to Fig.~\ref{fig:consistency-testing} for the different $S/N$ values adopted.}


\subsection{Precision}\label{sec:precision}

\begin{table*}
\centering
\caption{Accuracy and precision in \cosha{} for different levels of noise in GSL. For comparison we added the typical values for \citetalias{Yan2019}.}
\label{tab:prec-acc}
\begin{tabular}{lrrrrrrrrrrrrr}
\hline
 & \multicolumn{10}{c}{GSL} & \multicolumn{2}{c}{\citetalias{Yan2019}} \\
\hline
$S/N$ & \multicolumn{2}{c}{50} & \multicolumn{2}{c}{100} & \multicolumn{2}{c}{200} & \multicolumn{2}{c}{300} & \multicolumn{2}{c}{$\infty$}  \\
 & $\mu$ & $\sigma$ & $\mu$ & $\sigma$ & $\mu$ & $\sigma$ & $\mu$ & $\sigma$ & $\mu$ & $\sigma$ & $\mu$ & $\sigma$ \\
\hline
$T_\mathrm{eff}$~(K)       & $-$26.66 & 217.67 & $-$17.00 & 114.15 & $-$7.10  & 81.62 & $-$1.75 & 60.17 & $-$1.99 & 43.26 &  2.93 & 239.65 \\
$\log{g}$                  &   0.02 &   0.49 &  0.00  &   0.36 & $-$0.03 &  0.30  &    0.00   &  0.25 & 0.00  &  0.21 &  0.02 &   0.38 \\
$[\mathrm{Fe}/\mathrm{H}]$ &   0.05 &   0.55 &  0.03 &   0.41 &  0.02 &  0.32 & 0.02  &  0.26 & $-$0.01 &  0.16 & $-$0.02 &   0.24 \\
$[\alpha/\mathrm{Fe}]$     &   0.04 &   0.35 &  0.01 &   0.24 &  0.04 &  0.20  &  0.03  &  0.17 &  0.01 &  0.09 &  $-$0.01  &   0.08 \\
\hline
\end{tabular}
\end{table*}

\begin{figure*}
\centering
\includegraphics[width=0.9\textwidth]{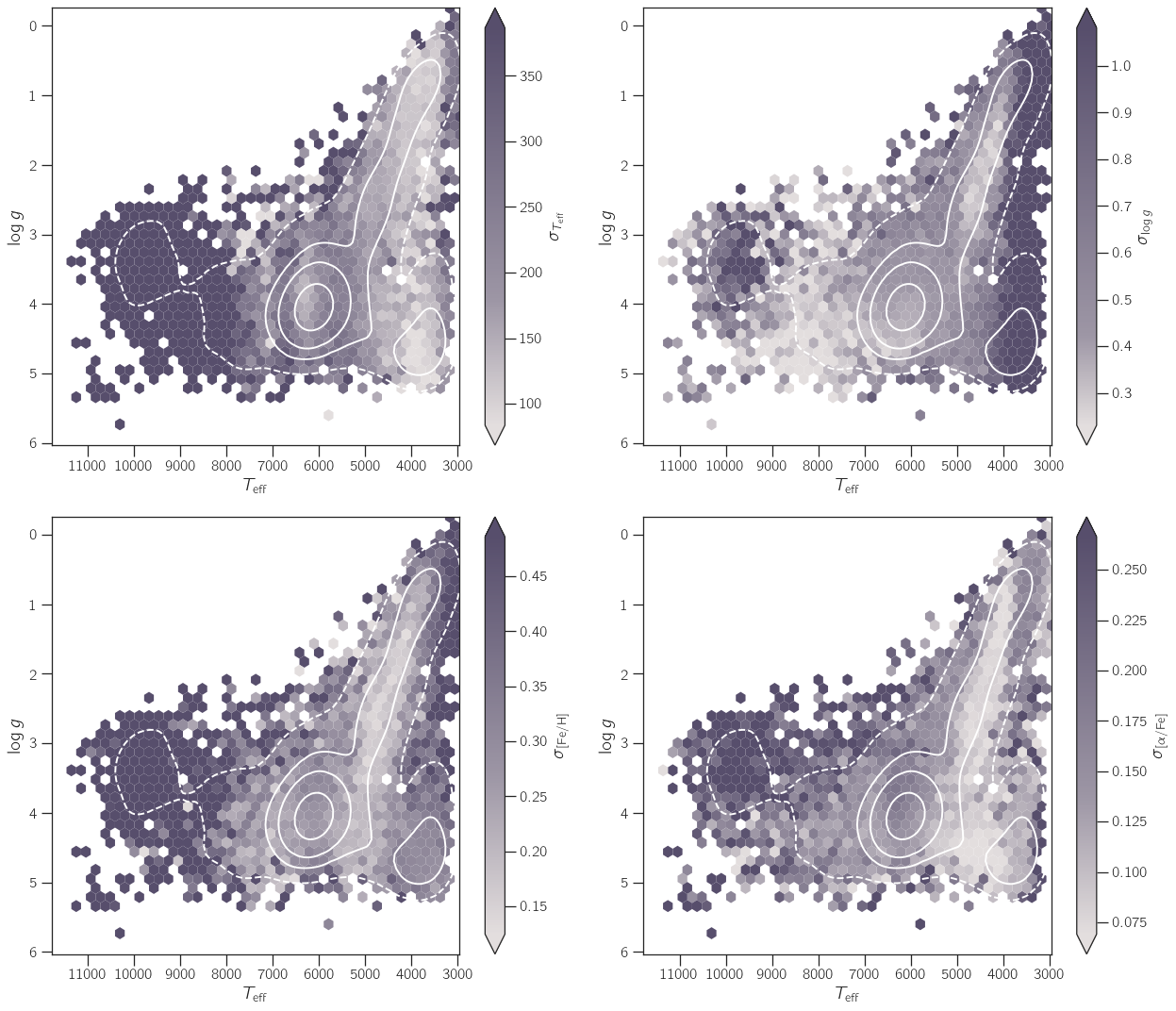}
\caption{Distribution of the internal precision of \cosha{} in each predicted property in the \logg{} vs. \teff{} plane. The contours represent the density distribution at the quartiles (solid) and percentile $95$th (dashed). The map of \teff{} precision (upper left) shows increasingly large imprecision towards higher temperatures. The \logg{} map reveals more imprecise results for the coolest stars. Both \FeH{} and \alphaM{} maps show increasing imprecision towards hotter and cooler stars. See text in \S~\ref{sec:precision} for details.}
\label{fig:logg-teff-precisions}
\end{figure*}
We quantify \cosha{} precision using the quantile predictions as outlined in \S~\ref{sec:cosha}. In Fig.~\ref{fig:logg-teff-precisions} we show the map of \cosha{} precision in the \logg{} \emph{versus} \teff{} plane. The contours represent the density distribution of stars. Clearly the loci of more precise determinations are not consistent with the locus of the highest star density. This seems to indicate that the origin of such (im)precision is not due to number statistics. The fact that most imprecise predictions correspond to temperature boundaries indicates that stars {at extreme values} of the parameter space are likely to have unreliable determinations. {This result proves one important limitation of ML (and arguably any stellar parameter estimation method): the determination of atmospheric properties depends on the existence of a comprehensive set of spectra (either theoretical or observed) with good quality stellar properties spanning an as wide as possible parameter space. From the distributions in Fig.~\ref{fig:logg-teff-precisions} we quote the following typical (median) precision, for each parameter in the cleaned $\sim22\,$k (and in the \citetalias{Yan2019}) sample: $\sigma_{\teff}\sim179(148)\,$K, $\sigma_{\logg}\sim0.42(0.42)$, $\sigma_{\FeH}\sim0.27(0.21)$ and $\sigma_{\alphaM}\sim0.14(0.09)$. When compared this numbers (in parenthesis) with those obtained from the $S/N$ simulations described above ($S/N=\infty$), we find that \cosha{} is actually underestimating its precision. On the other hand, when compared to the case $S/N=100$ (closer to the typical value for MaStar spectra) both estimates of the precision are closer to agreement, with \cosha{} still underestimating the precision for \teff{} and \logg{} but notably overestimating for the abundance parameters. Finally, we find that the determination of the precision by \cosha{} for the \citetalias{Yan2019} testing subset is generally congruent, with only \teff{} having an overestimated precision by $\sim60\,\%$.}

\subsection{Consistency with \citetalias{Yan2019}}\label{sec:mastardr1-consistency}

\figsetstart
\figsetnum{6}
\figsettitle{Consistency with GSL}

\figsetgrpstart
\figsetgrpnum{6.1}
\figsetgrptitle{Distribution of the discrepancies}
\figsetplot{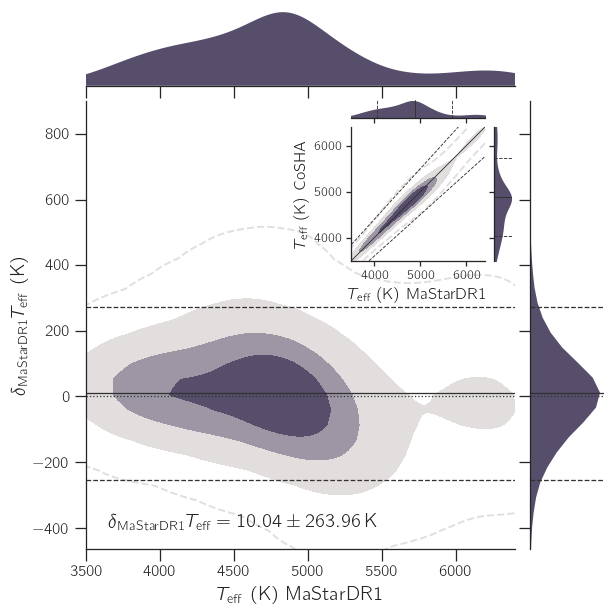}
\figsetplot{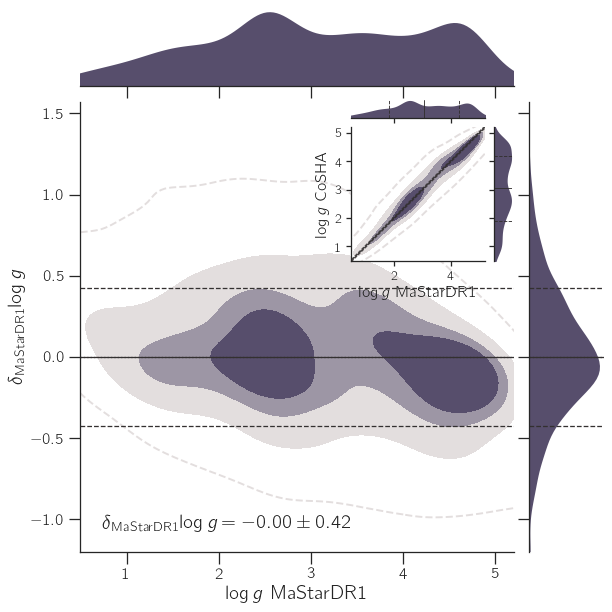}
\figsetplot{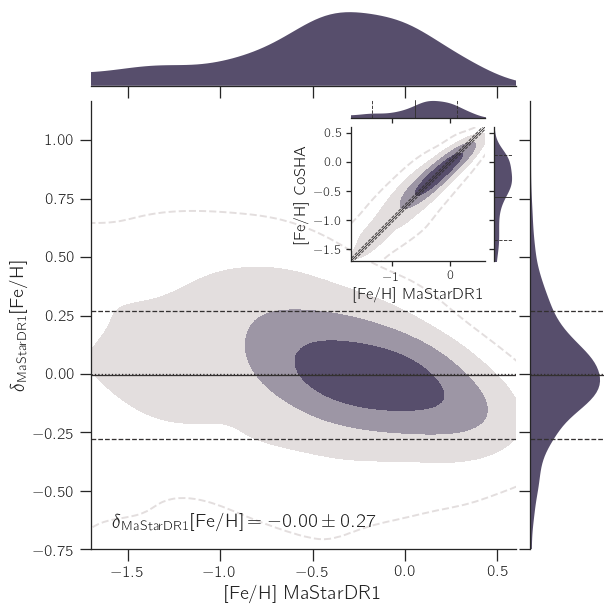}
\figsetplot{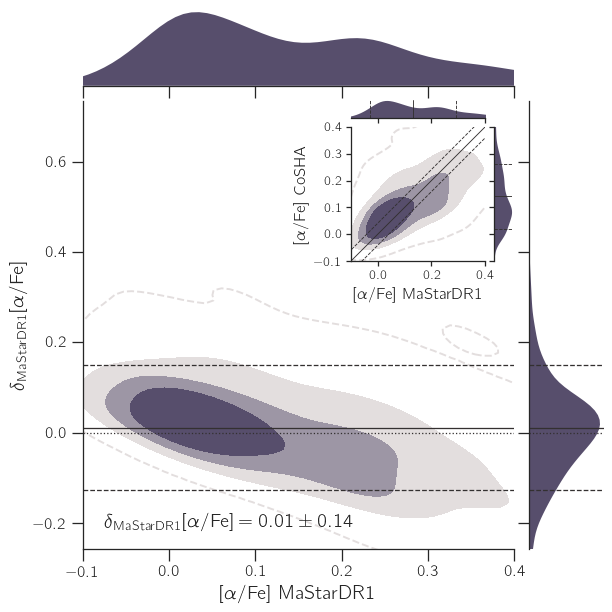}
\figsetgrpnote{Distribution of the discrepancies defined in Eq.~\eqref{eq:discrepancy} between \cosha{} and \citetalias{Yan2019} on the stellar properties \teff{}, \logg{}, \FeH{} and \alphaM{}. The color-coded contours represent the quartiles of the probability distribution. The dashed contours represent the $95\%$ confidence region to illustrate how spread are these distributions. The marginal distributions are also shown (purple) along with the corresponding mean (solid line) and standard deviation (dashed lines). The dotted grey line represents a perfect consistency ($\delta \lab{}=0$). A direct comparison between the stellar properties \teff{}, \logg{}, \FeH{} and \alphaM{} retrieved in this study and \citetalias{Yan2019} is also show in the inset axes.
See text in \S~\ref{sec:mastardr1-consistency} for details.}
\figsetgrpend

\begin{figure*}
\centering
\includegraphics[width=0.45\textwidth]{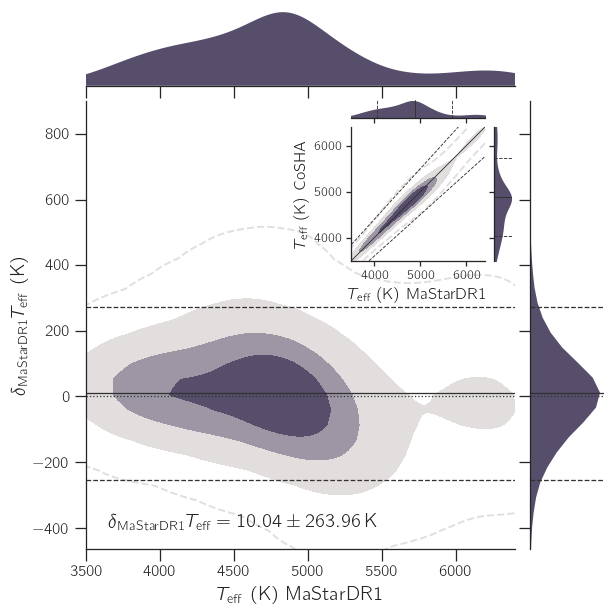}
\includegraphics[width=0.45\textwidth]{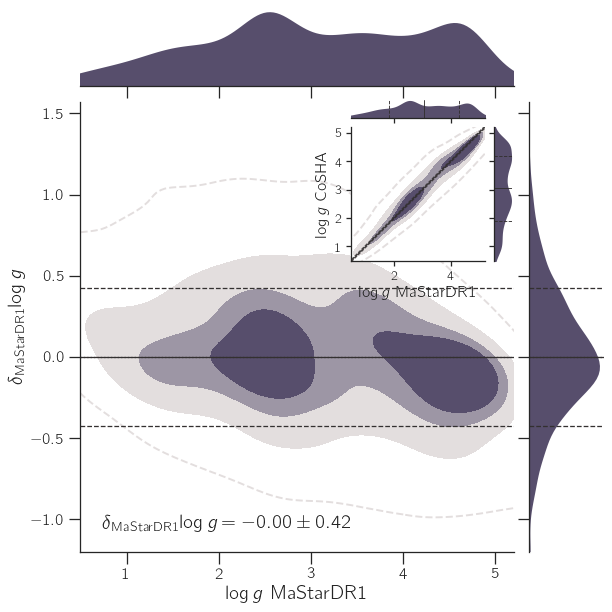}
\includegraphics[width=0.45\textwidth]{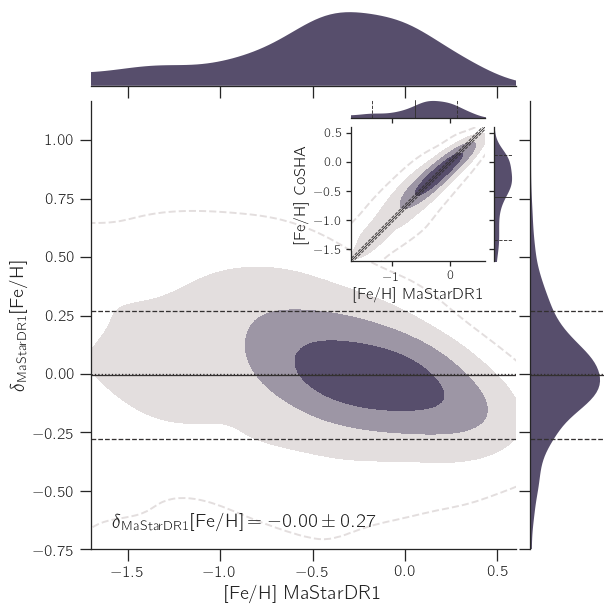}
\includegraphics[width=0.45\textwidth]{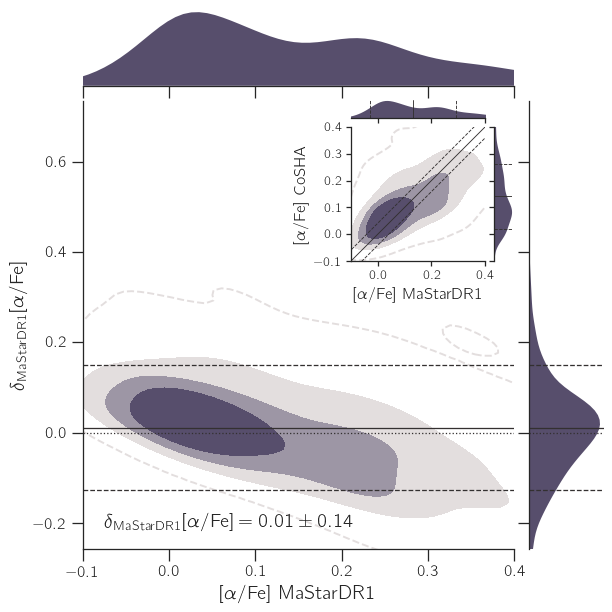}
\caption{Distribution of the discrepancies defined in Eq.~\eqref{eq:discrepancy} between \cosha{} and \citetalias{Yan2019} on the stellar properties \teff{}, \logg{}, \FeH{} and \alphaM{}. The color-coded contours represent the quartiles of the probability distribution. The dashed contours represent the $95\%$ confidence region to illustrate how spread are these distributions. The marginal distributions are also shown (purple) along with the corresponding mean (solid line) and standard deviation (dashed lines). The dotted grey line represents a perfect consistency ($\delta \lab{}=0$). A direct comparison between the stellar properties \teff{}, \logg{}, \FeH{} and \alphaM{} retrieved in this study and \citetalias{Yan2019} is also show in the inset axes.
See text in \S~\ref{sec:mastardr1-consistency} for details. The complete figure set (6 images) is available in the online journal and includes images that show the consistency with GSL, CANNON, Gaia, \cosha{}, and IndoUS.
}
\label{fig:consistency-mastardr1}
\end{figure*}

\figsetgrpstart
\figsetgrpnum{6.2}
\figsetgrptitle{Consistency with GSL}
\figsetplot{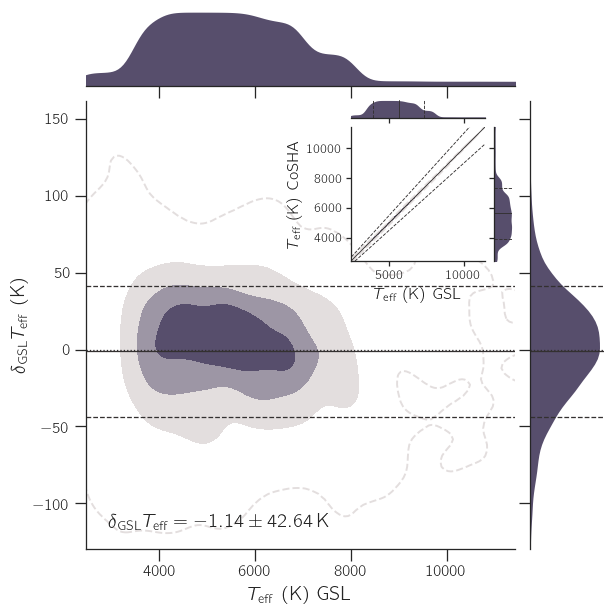}
\figsetplot{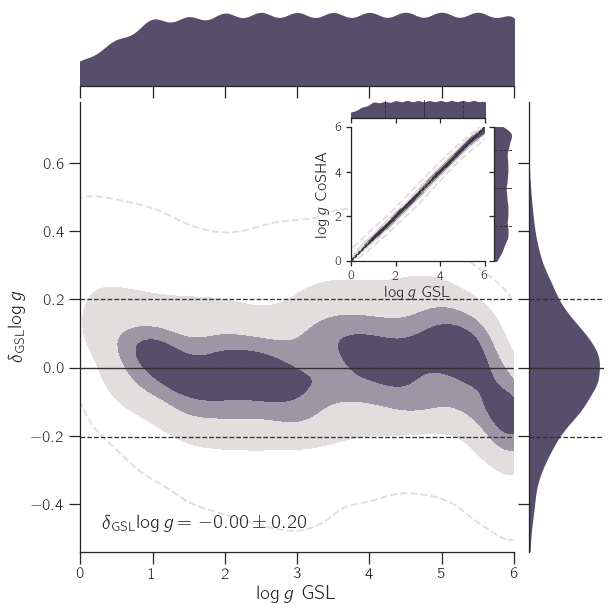}
\figsetplot{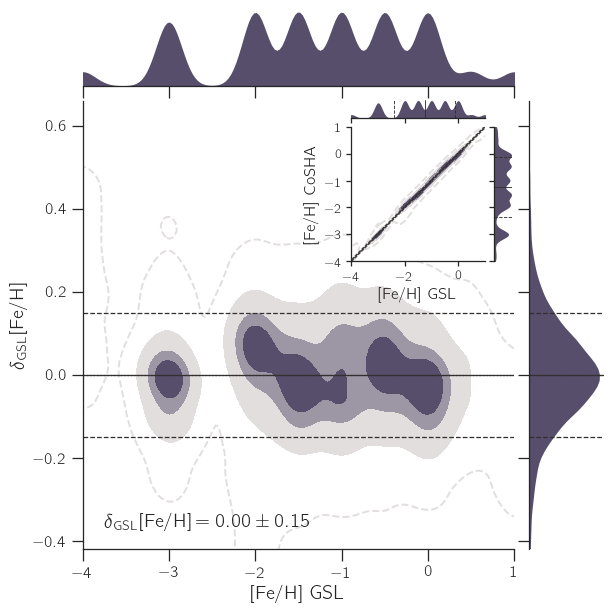}
\figsetplot{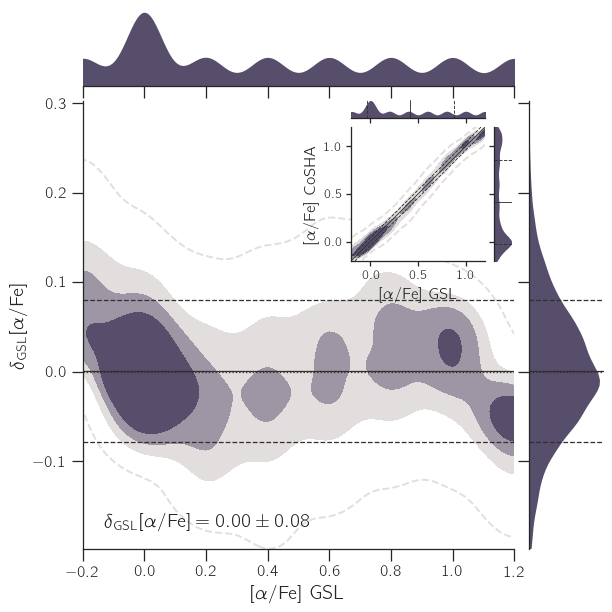}
\figsetgrpnote{Similar to Fig.~\ref{fig:consistency-mastardr1} but comparing our estimates for GSL the true values. As expected the discrepancies with the GSL are notoriously smaller if compared to \citetalias{Yan2019} and APOGEE-ASPCAP/CANNON.}
\figsetgrpend

\figsetgrpstart
\figsetgrpnum{6.3}
\figsetgrptitle{Consistency with CANNON}
\figsetplot{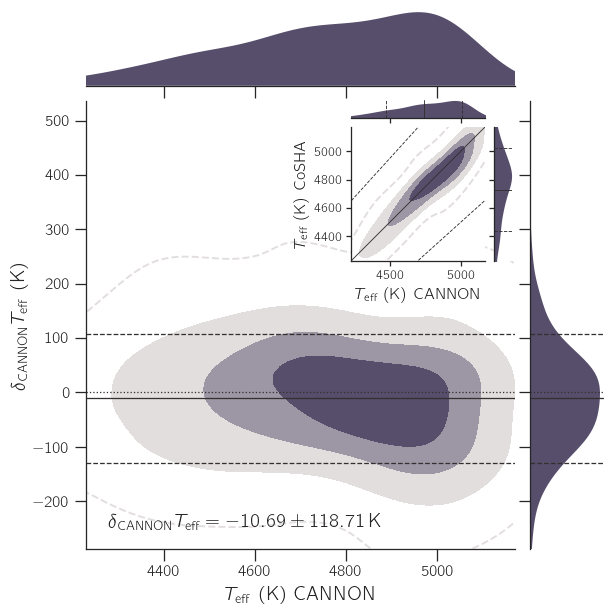}
\figsetplot{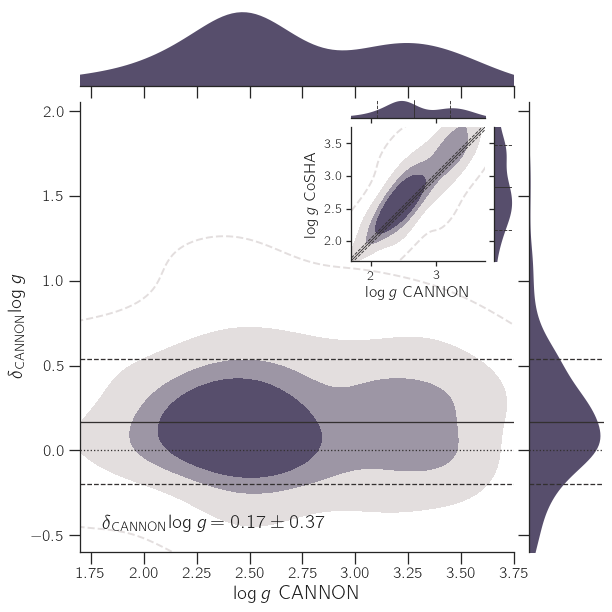}
\figsetplot{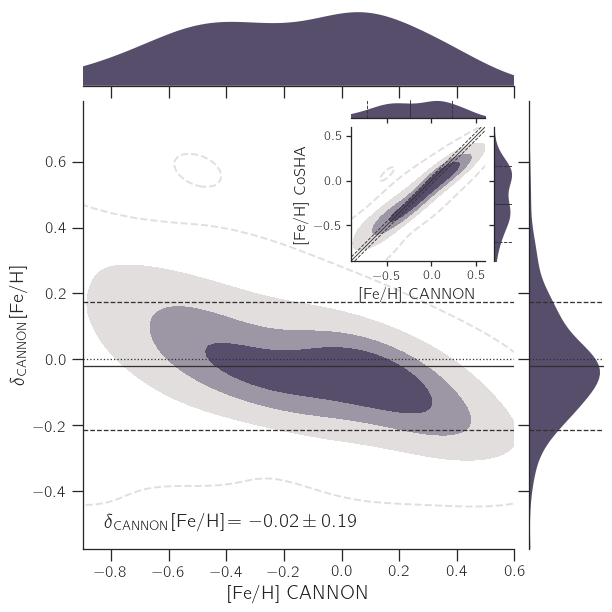}
\figsetplot{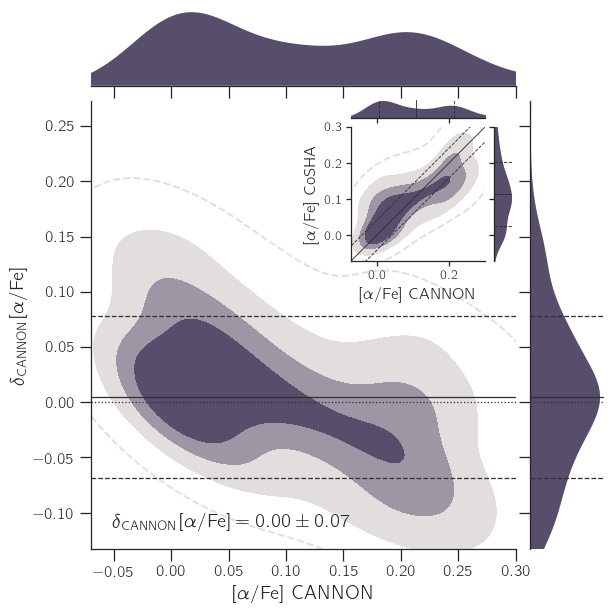}
\figsetgrpnote{Similar to Fig.~\ref{fig:consistency-aspcap} but comparing with the APOGEE-CANNON estimates. The consistency between the APOGEE-CANNON and \cosha{} estimates is statistically indistinguishable from that of APOGEE-ASPCAP.}
\figsetgrpend

\figsetgrpstart
\figsetgrpnum{6.4}
\figsetgrptitle{Consistency with Gaia}
\figsetplot{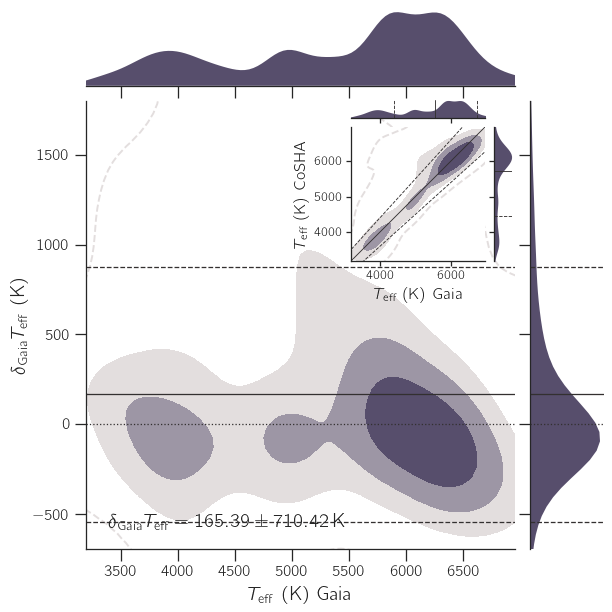}
\figsetplot{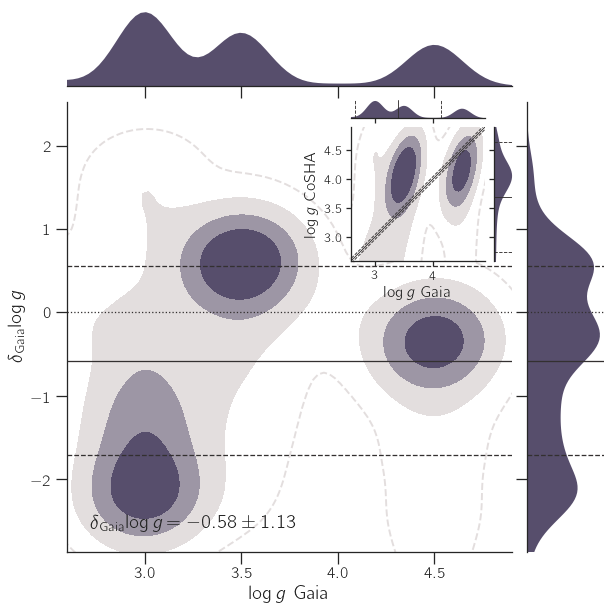}
\figsetplot{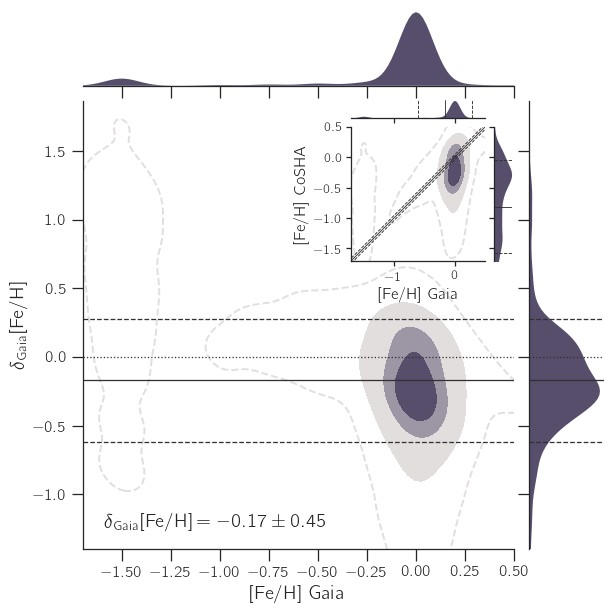}
\figsetgrpnote{Similar to Fig.~\ref{fig:consistency-mastardr1} but comparing with the Gaia estimates. \teff{} is the most consistent property, with most of the sample within $\pm10\%$ consistency. \logg{} and \FeH{} on the other hand, show large inconsistencies, specially the former property. We should note though, that Gaia photometric system is not suitable to constrain these properties to accuracy comparable to spectroscopic data.}
\figsetgrpend

\figsetgrpstart
\figsetgrpnum{6.5}
\figsetgrptitle{Consistency with \cosha{}}
\figsetplot{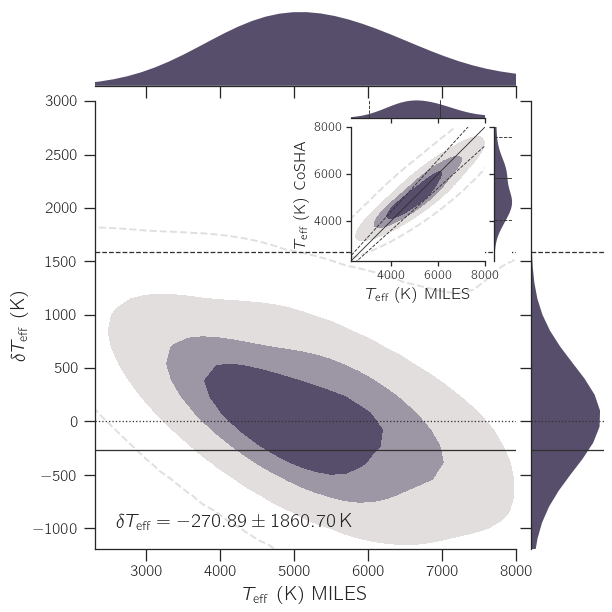}
\figsetplot{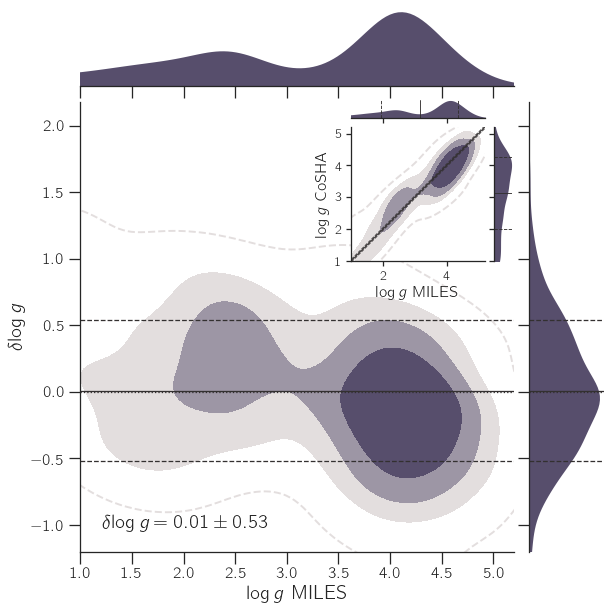}
\figsetplot{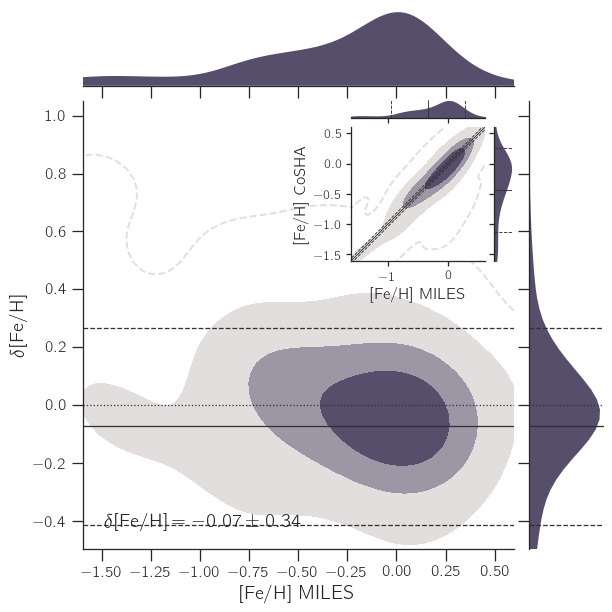}
\figsetgrpnote{Similar to Fig.~\ref{fig:consistency-mastardr1} but comparing \cosha{} estimates for the MILES spectra and those distributed with the library.}
\figsetgrpend

\figsetgrpstart
\figsetgrpnum{6.6}
\figsetgrptitle{Consistency with IndoUS}
\figsetplot{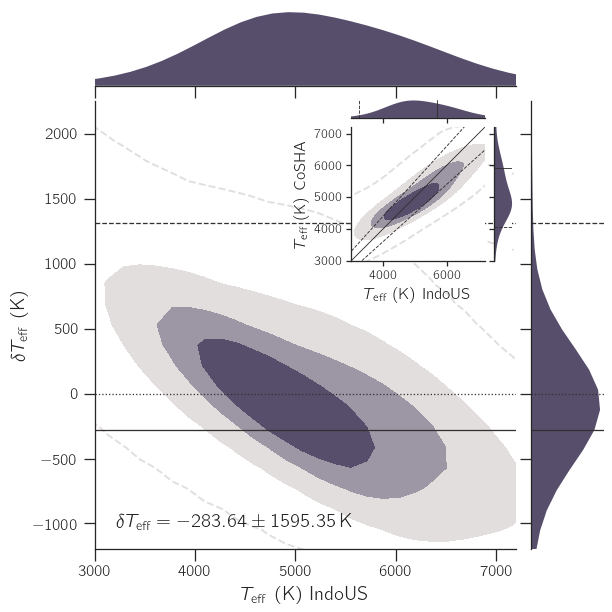}
\figsetplot{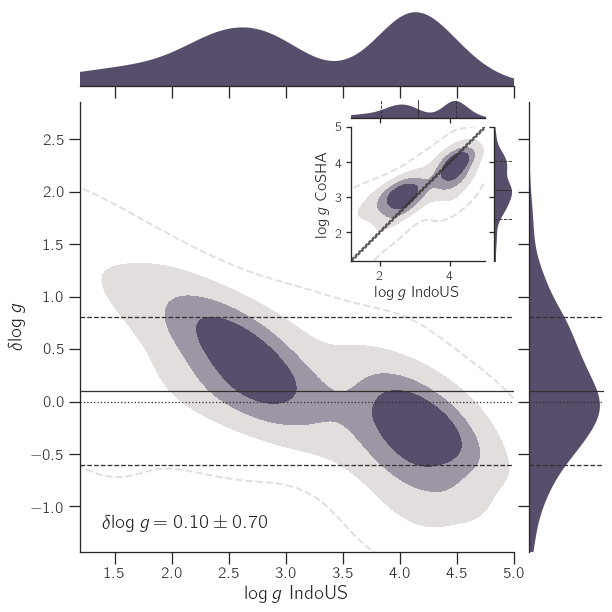}
\figsetplot{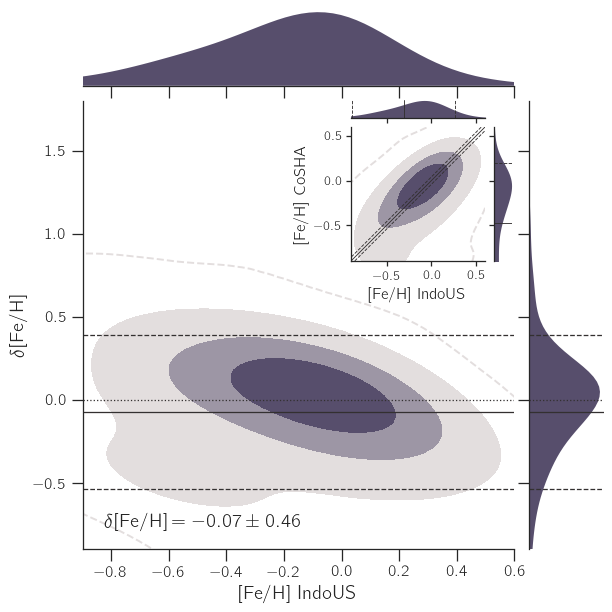}
\figsetgrpnote{Similar to Fig.~\ref{fig:consistency-miles} but for IndoUS. \teff{} and \logg{} are the most inconsistent parameters, showing trends that suggest a limitation in extreme values of the parameter space for \cosha{}.}
\figsetgrpend

\figsetend

Fig.~\ref{fig:consistency-mastardr1} shows the discrepancy in \teff{}, \logg{}, \FeH{} and \alphaM{}, as defined in Eq.~\ref{eq:discrepancy}, between the values estimated in this study and those reported in \citetalias{Yan2019}. We only compare those stars for which \citetalias{Yan2019} made estimations in our clean sample ($\sim3\,$k stars).
It is clear from these comparisons that \teff{} is the most robust property (i.~e. less independent on the methodology), having the best consistency ($<10\%$ discrepancy) across the whole range. Both the mean and the standard deviation in the marginal distributions are in agreement within $\sim2\,$K and $\sim22\,$K for \teff{}, and $\sim0.03$ and $\sim0.09\,$dex for the rest of the properties, respectively. Since most of these stars belong to the training set ($\sim90\%$), we expect an overall high consistency in this particular comparison. However, it is interesting to note that some trends may appear given the different methodologies adopted by \citetalias{Yan2019} and in this study. We note that this discrepancies are comparable to the \citetalias{Yan2019} instrinsic errors estimated in the previous section. This indicates that even though we used the \citetalias{Yan2019} spectra to train \cosha{}, we are able to disentangle the typical (intrinsic) error of both methods above. Furthermore, the observed discrepancies in Fig.~\ref{fig:consistency-mastardr1} are likely to be dominated by \citetalias{Yan2019} errors.

\teff{} and \logg{} discrepancies show almost no deviation from the Gaussian distribution. However, \FeH{} and \alphaM{} display clear trends with respect to the \citetalias{Yan2019} estimates. In the particular case of the \alphaM{}, the distribution of discrepancies displays a negative slope with respect to the values reported by \citetalias{Yan2019}: the higher the \alphaM{}, the more inconsistent becomes our estimate with respect to \citetalias{Yan2019} estimate, in the negative sense. This inconsistency may originate in either method or (most likely) in both. Therefore we will need to compare to external estimates of these parameters in order to find clues on the origin of this trend.

\subsection{Consistency with APOGEE}\label{sec:apogee-consistency}

\begin{figure*}
\centering
\includegraphics[width=0.45\textwidth]{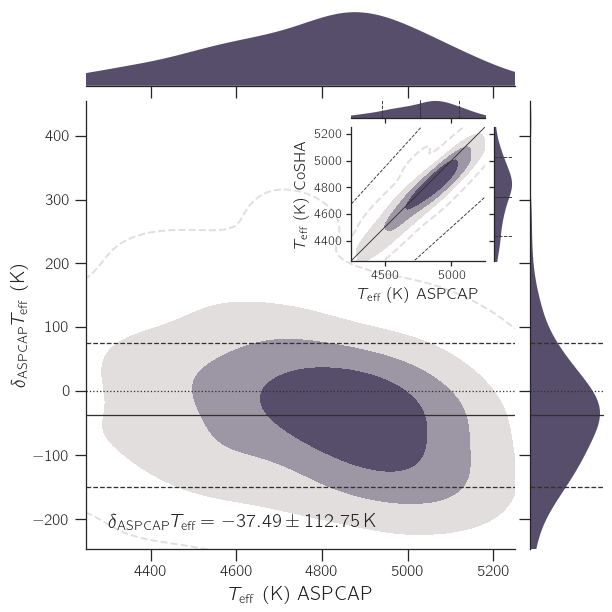}
\includegraphics[width=0.45\textwidth]{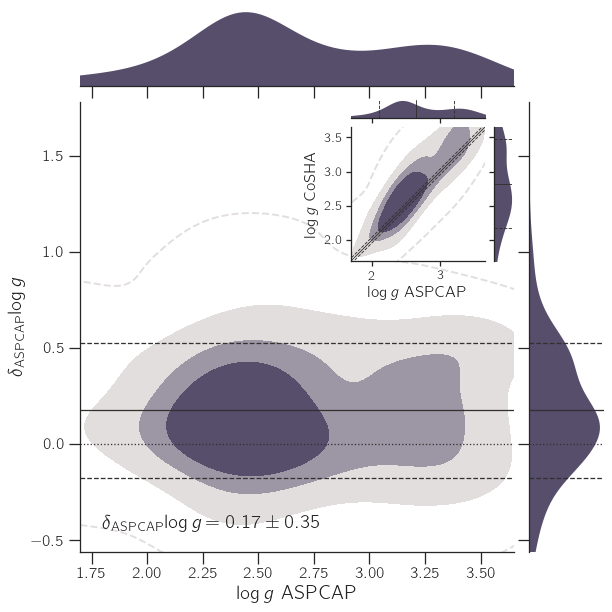}
\includegraphics[width=0.45\textwidth]{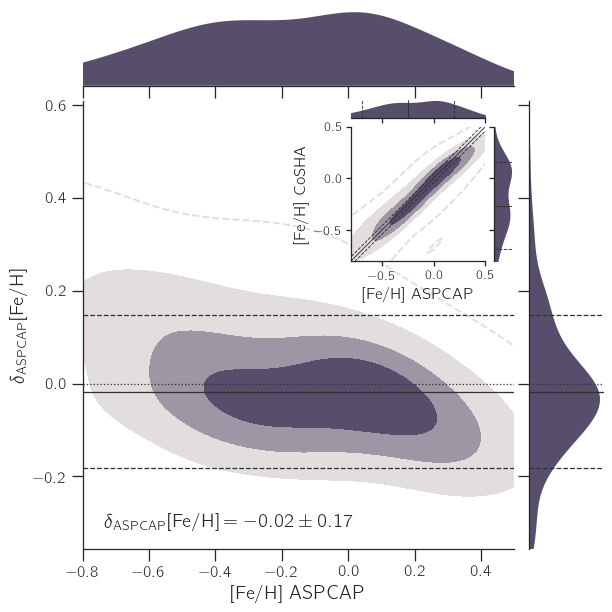}
\includegraphics[width=0.45\textwidth]{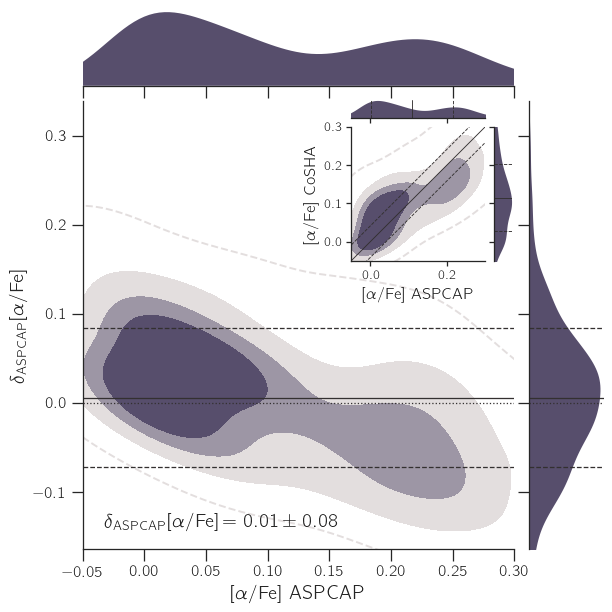}
\caption{Similar to Fig.~\ref{fig:consistency-mastardr1} but comparing our estimates to those derived by the APOGEE spectral analysis pipeline (ASPCAP). The systematic discrepancies with APOGEE are larger than those found with \citetalias{Yan2019}. In summary, \cosha{} predicts systematically cooler and more dwarf stars and marginally Fe-poorer than APOGEE, with essentially no bias in the \alphaM{} estimates. See text in \S~\ref{sec:apogee-consistency} for details.}
\label{fig:consistency-aspcap}
\end{figure*}
The APOGEE survey offers a great validation tool since it is known to have accurate distributions of \teff{}, \logg{} and abundances of several chemical species, including $\alpha$-elements and Fe \citep[e.~g.,][]{Jonsson2018}. The APOGEE DR14 reports \teff{}, \logg{}, \FeH{} and \alphaM{} values for most of the stars in the survey using two different methods: ASPCAP \citep{GarciaPerez2016} and the CANNON \citep{Ness2015}. In the following we compare only with the original method developed for the APOGEE: ASPCAP, but see the supplementary material. Since the target selection for the MaStar survey results from the combination of piggybacking on the APOGEE plates and cherry-picking to improve the sampling of the parameter space (specially in the \teff{}), some stars in our clean sample of MaStar belong to the APOGEE survey, by construction. However, the sample cleaning described in \S~\ref{sec:train-test-sets} and the match with Gaia to retrieve distances, cut the matching subset between MaStar and APOGEE down to just $\sim400$ stars. In Fig.~\ref{fig:consistency-aspcap} we show the comparison between the physical properties derived by \cosha{} and those published by APOGEE for those stars in common in both surveys. As in the previous sections, we find that the \teff{} determinations are the most consistent results, with most of the stars within the boundary of a $10\%$ discrepancy (c.~f. \teff{} inset plot). Overall, we report a discrepancy with APOGEE in \teff{}, in the mean and the standard deviation of the marginal distributions of $\sim-45$ and $\sim101\,$K, respectively. The distribution of \logg{} show the larger scatter around the perfect consistency line. However the overall shape of the marginal distributions are rather consistent. These distributions disagree in their mean and standard deviation by at most $\sim0.18$ and $\sim0.35\,$dex, respectively. The marginal distributions of \FeH{} and \alphaM{}, on the other hand, show smaller discrepancies in their mean and standard deviation of at most $\sim-0.01$ and $0.15\,$dex, respectively, \alphaM{} being strikingly consistent.

The discrepancies in \teff{} and \logg{} show no trend with respect to the APOGEE estimate. Notwithstanding, both \teff{} and \logg{} show in fact a systematic discrepancy, whereby \cosha{} seems to over-predict cooler and more dwarf stars than APOGEE. We expect the APOGEE error estimations to account for some of the observed discrepancy ($\Delta\teff\sim79\,$K and $\Delta\logg\sim0.05\,$, respectively). However, the remaining discrepancy is likely to have an origin in \cosha{}. \FeH{} shows a mild discrepancy systematic and almost no trend with the values reported by APOGEE. \alphaM{}, on the other hand, shows a trend between its discrepancy and the values reported for APOGEE: the higher the APOGEE estimates, the larger the discrepancy in the negative sense. It is worth mentioning that a similar trend was observed in the discrepancies with \citetalias{Yan2019}. It is encouraging though, that the mean and standard deviation of the distribution of discrepancies are overall rather small.

{The observed trends in the abundance parameters when comparing \cosha{} to APOGEE and \citetalias{Yan2019} (largely based on APOGEE) have been reported before by other authors using different methods \citep[e.~g.,][]{Ting2019, Nandakumar2020}. \citet{Ting2019} in their Figs.~11 show similar trends for O, Mn, Ca, Ti. \citet{Nandakumar2020} in their Fig.~1 show similar trends for \alphaM{} (based on \textsc{cannon}). Interestingly, when training (\textsc{cannon}) using the APOGEE labels (their Fig.~2) those trends do not appear or are, at least, mitigated. None of these groups of author explain these trends in detail. However the fact that such trends are stronger when training with labels different than those predicted for APOGEE may indicate a mismatch between APOGEE labels and the spectra predicted for those labels using generative models as The Payne and \textsc{cannon}. As a matter of fact, theoretical recipes for stellar atmospheres predictions are known to make \emph{ad hoc} assumptions that may turn into inconsistencies in the output stellar spectrum. We recall that CoSHA (GSL), APOGEE/MaStar \citep[ATLAS9,][]{Meszaros2012} are based on different theoretical libraries to make physical property predictions. \citet{Lancon2021}, for example, compared GSL with X-Shooter stellar library and found several discrepancies across the HR diagram. They argue that such inconsistencies may originate from the use of different theoretical prescriptions.}

\section{The parameter space distribution}\label{sec:parameter-distributions}

\subsection{a partial volume correction for MaStar}\label{sec:volume-correction}

\begin{figure*}
\centering
\includegraphics[width=0.9\textwidth]{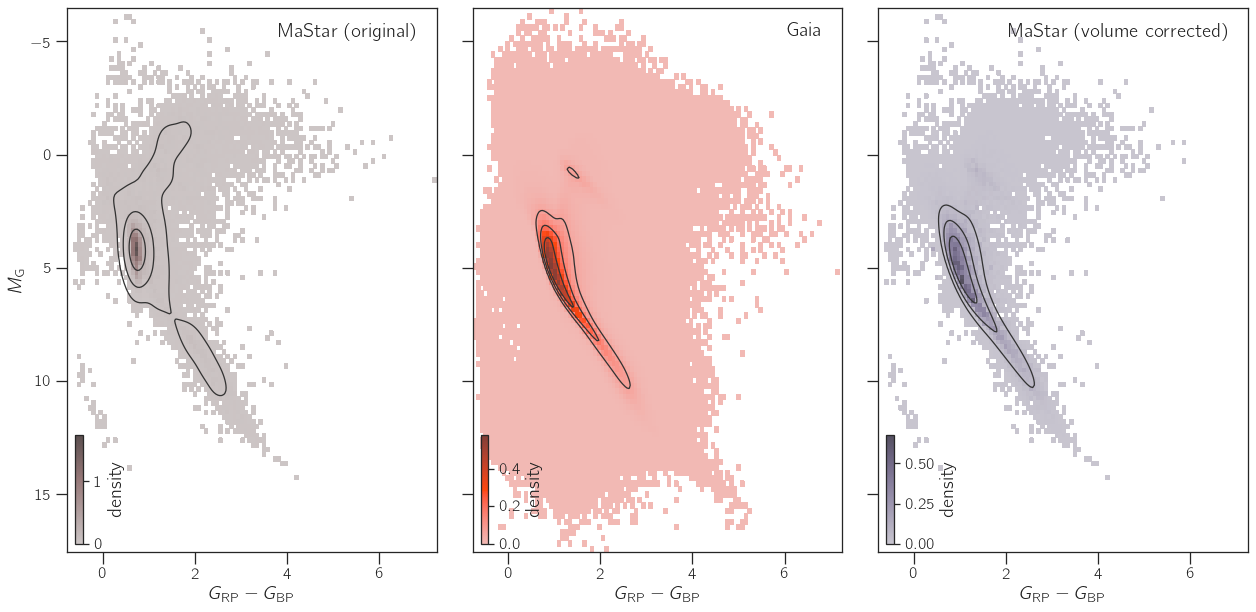}
\caption{CMD of the MaStar library (left), Gaia DR2 (middle) and the volume corrected version of MaStar (right) are shown. The similarity in the shape of the middle and right distributions shows that the volume correction is accurate.}
\label{fig:volume-correction}
\end{figure*}
The MaStar library is intended to provide a homogeneous and complete coverage of the parameter space in \teff{}, \logg{}, \FeH{}, and \alphaM{}. Hence, by construction, {the corresponding physical properties are not representative of the distribution of stellar populations in the Milky Way, i.~e. intrinsically rare stars may be over-represented.} In order to reliably compare the distributions of these physical properties with those already known from other (volume-complete) surveys, we first need to make a volume correction. In principle, we would need a sample of the stars representative of all the Milky Way stellar populations. However, to the best of our knowledge such data set does not exist. Our best choice to date is to use the Gaia survey DR2 \citep{Brown2018}, containing well over $60\,$M unique stars and complete down to $12\,$mag in the $G$ band. Since the Gaia survey is not volume-complete and the MaStar sample is not representative of all plausible stars, this is a \emph{partial} volume correction \citep{Bailer-Jones2018}. Therefore, the volume corrected MaStar sample is only representative of the stellar populations (as seen in the color-magnitude distribution) sampled by the Gaia survey \citep{Evans2018, Arenou2018}.


We compute the volume correction using the distribution of stars in the color-magnitude diagram (CMD). Mathematically, we express the volume as:
\begin{equation}\label{eq:volume-correction}
V\equiv\frac{\text{PDF}_{\text{MaStar}}}{\text{PDF}_{\text{Gaia}}},
\end{equation}
where the $\text{PDF}_{\text{MaStar}}$ and $\text{PDF}_{\text{Gaia}}$ are the corresponding probability density distributions of observing stars in the CMD for the corresponding catalog. {We implement a KDE to estimate these PDFs from the MaStar and Gaia samples. We download the Gaia source catalogs from the archives\footnote{\url{http://cdn.gea.esac.esa.int/Gaia/gdr2/gaia_source/csv/}.} and applied the cuts documented in \citet[][\S~2]{Arenou2018} in order to build the H-R diagram. To correct the CMD from the MW extinction effects, we followed \citet[][and references therein]{Alzate2021}. The absolute magnitude, $M_G$ was calculated using the relation in \citet{Arenou2018} from the parallax estimation. After applying these cuts we have a sample of $\sim66\,$M stars from Gaia.} In Fig.~\ref{fig:volume-correction} we show the original distribution in the MaStar (left), the resulting Gaia distribution (middle) and the corrected MaStar distribution (right).

\subsection{The MaStar properties distributions}\label{sec:mastar-parameters}

\begin{figure*}
\centering
\includegraphics[width=0.9\textwidth]{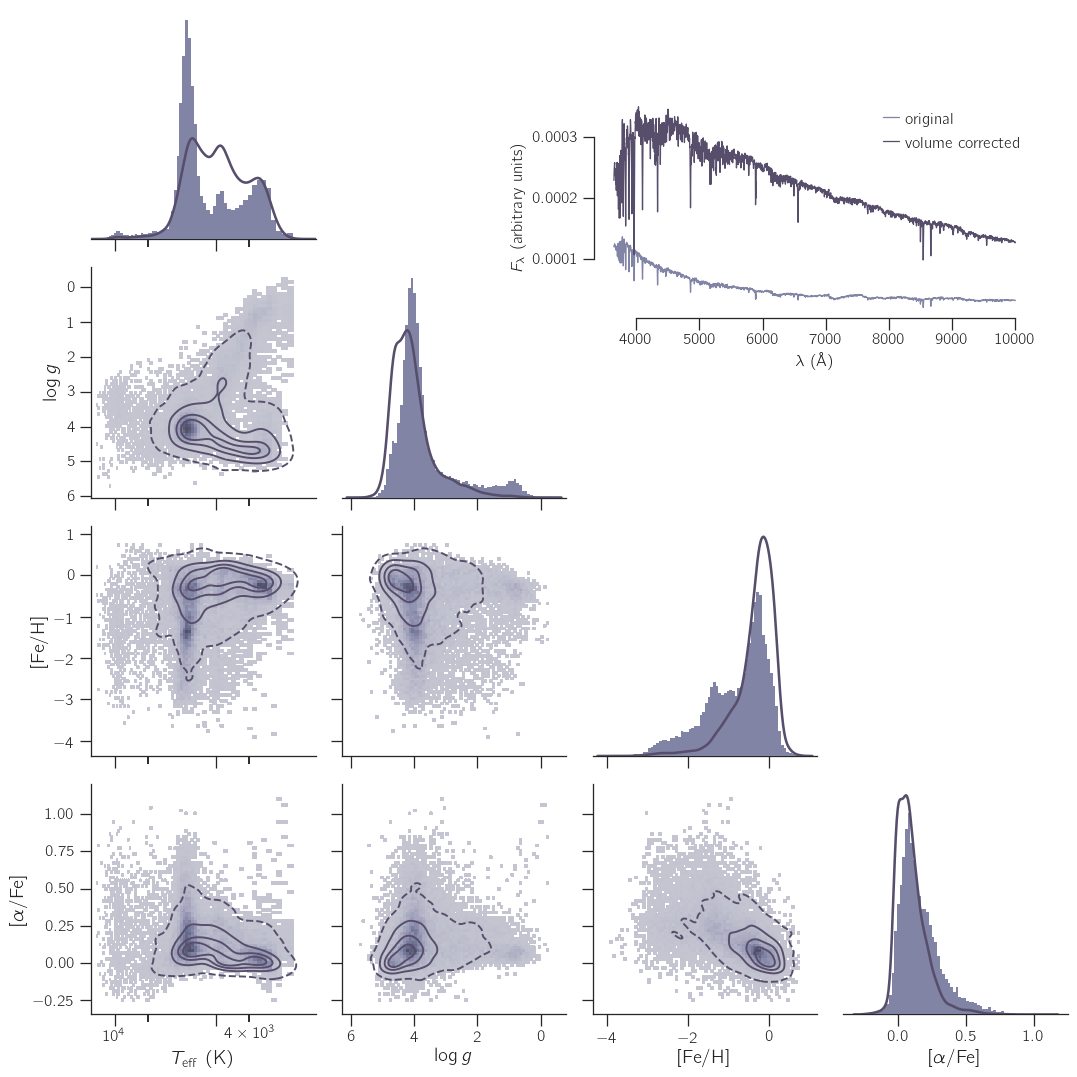}
\caption{Similar to Fig.~\ref{fig:training-testing-sets}, the joint distribution of the stellar parameters for MaStar as estimated by \cosha{}. In light and dark purple we show the original and the volume corrected distributions as explained in \S~\ref{sec:volume-correction}, respectively. The contours are as in Fig.~\ref{fig:logg-teff-precisions}. The corresponding average spectrum of the MaStar library and the volume corrected one (inset axes). The in-homogeneity in the MaStar sampling of the parameter space become apparent. \textit{(i)} \teff{} becomes flatter after correction; \textit{(ii)} \logg{} is corrected for under-sampling of giant stars and; \textit{(iii)} \FeH{} and \alphaM{} become more biased towards solar abundance patterns. See text in \S~\ref{sec:parameter-distributions} for details.}
\label{fig:parameter-distributions}
\end{figure*}
In the previous sections we explored the MaStar distribution of the stellar properties through its marginal distributions (c.~f. Figs.~\ref{fig:consistency-mastardr1} and \ref{fig:consistency-aspcap}). In Fig.~\ref{fig:parameter-distributions} we introduce the final joint distribution of the stellar properties for the cleaned version of the (whole; $\sim22\,$k) MaStar library. The light purple distributions represent the raw estimates from our model. Since by construction these distributions are not derived for a representative sample (see Fig.~\ref{fig:volume-correction}), this means that we cannot confidently validate our results by comparing with the currently known trends in the Milky Way (MW) provided by APOGEE \citep[e.~g.,][]{Hayden2015}. In order to solve for this issue, we implement the (partial) volume correction described in \S~\ref{sec:volume-correction}.

In Fig.~\ref{fig:parameter-distributions} we show the volume corrected sample of MaStar parameters in dark purple. Interestingly, \teff{} becomes flatter after correction, thus the peak around $\teff\sim6000\,$K becomes less pronounced in favor of cooler stars. In the \logg{} distribution dwarf stars ($\logg\sim5\,$) become more relevant. The \FeH{} distribution in the raw MaStar distributions shows an oversampling of low iron abundance ($\FeH\lesssim-1\,$) that is redistributed towards solar abundance after volume correction. \alphaM{} is perhaps the most unchanged distribution with a slight redistribution of $\alphaM\sim0.5$ towards solar values. It is worth noticing that well-known distributions such as the \logg{} \emph{versus} \teff{} and the \alphaM{} \emph{versus} \FeH{} seem to follow the expected trends: relatively cool dwarf (main sequence) stars dominate the distributions, with abundances biased towards solar values \citep[e.~g.,][]{Hayden2020}. {More interesting would be to see if we can recover the spatial trends in the abundances subspace.}

\subsection{Spatial distributions}\label{sec:spatial-distribution}

\begin{figure*}
\centering
\includegraphics[width=0.9\textwidth]{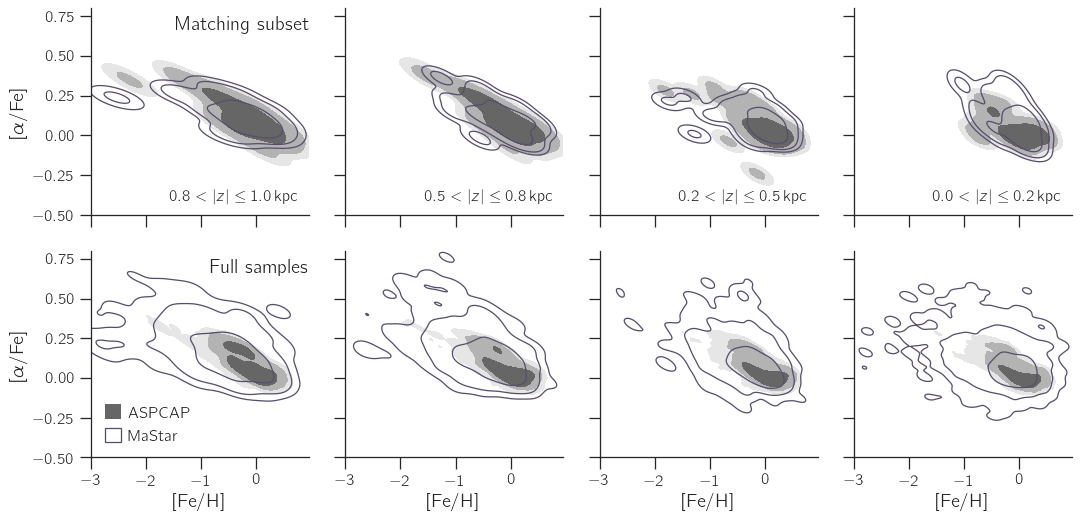}
\caption{Spatial distribution of the MaStar stars (purple) in the \alphaM{} vs. \FeH{} plane, segregated by the vertical distance to the MW disk plane, $z$. As a reference we use the corresponding distribution for APOGEE-ASPCAP (grey). In the top panels we compare only the matched subset of stars between the two surveys and in the bottom panels we compare the trends found in the full samples for both surveys. In the top panels it is appreciated that MaStar and APOGEE are statistically indistinguishable from each other, for any vertical distance. The full samples are still compliant, albeit the larger scatter in MaStar, with the expected trends: stars become more Fe-poor and $\alpha$-enhanced towards the thick disk, whilst in the thin disk stars with solar abundances dominate. See text in \S~\ref{sec:spatial-distribution} for details.}
\label{fig:alpha-fe-spatial}
\end{figure*}

{We set out to investigate the spatial trends in the \alphaM{} versus \FeH{} plane resulting from CoSHA (on MaStar spectra) and compare to APOGEE. We match the APOGEE sample ($\sim250\,$k stars) with Gaia DR2 in order to have distances of each star in APOGEE. Then we match the resulting APOGEE subset with estimated distances ($\sim64\,$k) to the MaStar sample, resulting in $\sim400$ stars}. In Fig.~\ref{fig:alpha-fe-spatial} we show the spatial distribution in the $z$ direction of the galactic plane, for the stars in MaStar (purple) and the APOGEE (grey) surveys. In the top panels we show the matched subset of stars between both surveys (APOGEE and MaStar). In all the panels the expected trends appear whereby stars become more Fe-poor and $\alpha$-enhanced towards the thick disk (higher $|z|$), and they present abundances more similar to the solar one towards the thin disk \citep[e.~g.,][]{Hayden2015, Nandakumar2020}. In the bottom panels we compare the full samples to see if the aforementioned trends remain in MaStar. Although the scatter is clearly larger in this case, the trends are still significant. It is interesting that we retrieve these trends using \cosha{} since, in principle, the training process is unaware of the spatial distribution of the stars. This result reinforces the robustness of our \cosha{}, already drawn from the internal and external tests (c.~f., \S~\ref{sec:model-evaluation}).


\section{Summary and conclusions}\label{sec:conclusions}

We present the novel Code for Stellar properties Heuristic Assignment (\cosha{}) to estimate atmospheric parameters from the MaStar stellar optical spectral energy distributions, namely: \teff{}, \logg{}, \FeH{} and \alphaM{}. \cosha{} implements a {conventional} machine learning (ML) approach named Gradient Tree Boosting (GTB) which consists in training a predefined number of decision trees sequentially, each improving its predecessor. {Because the code base of \cosha{} is small (once the training sample is cleaned), the main strength of our approach is that it is easier implement, interpret its results (as is based in decision trees) and easier to scale than more commonly used approaches such as ANN, as there are less hyper-parameters to tune \citep[e.~g.][]{Geron2017, Chollet2017, Ivezic2019}. Once trained, \cosha{} only requires the input spectra to be in the same spectral/parameter space as the training sample to predict reliable atmospheric properties}.

For the training and testing samples we used a combination of empirical (\citetalias{Yan2019}) and theoretical (GSL) libraries. Both have advantages and disadvantages that were explained (see \S~\ref{sec:train-test-sets}). {Despite of those we were able to use train a model that yields reliable results. Based on the internal tests (i.~e., comparing the input with the output) using the testing subset, an overall performance of \cosha{} of: $\Delta\teff\sim-1.4\pm90.0\,$K, $\Delta\logg\sim0.002\pm0.246$, $\Delta\FeH\sim-0.004\pm0.174$ and $\alphaM\sim0.004\pm0.088$. Moreover, the performance of \cosha{} on the segregated \citetalias{Yan2019} and GSL subsets showed a systematically more imprecise estimate of the physical properties on the former, even after the internal errors from \cosha{} were removed. The fact that the errors on the GSL properties are smaller than in the \citetalias{Yan2019} points towards an origin in the \citetalias{Yan2019} subset itself. We reckon several possibilities: instrumental systematics and random noise and and/or internal errors introduced by the different methods adopted by \citetalias{Yan2019}. None of these issues are present in the GSL subset. When introducing random noise in GSL spectra, we are able to reconcile these differences.} The main results can be summarized as follows:

\begin{itemize}
    \item We found no statistically significant difference on the distributions of the predicted parameters between the values derived by \cosha{} and those published by \citetalias{Yan2019}. The systematic and random discrepancy between such distributions are: $\delta\teff\sim10\pm264\,$K, $\delta\logg\sim0.00\pm0.42$, $\delta\FeH\sim-0.00\pm0.27$ and $\delta\alphaM\sim0.01\pm0.14$. Most of the parameters show no trend between the discrepancy and the reference \citetalias{Yan2019}'s values, except for \alphaM{}. This parameter becomes increasingly more inconsistent (with respect to \citetalias{Yan2019}) for higher values of \alphaM{}.
    
    \item The comparison between \cosha{} and APOGEE-ASPCAP estimates on a subset of ($\sim400$) stars in common between both surveys, revealed systematic discrepancies comparable to those in the previous item: $\delta\teff\sim-37\pm113\,$K, $\delta\logg\sim0.17\pm0.35$, $\delta\FeH\sim-0.02\pm0.17$ and $\delta\alphaM\sim0.01\pm0.08$. The errors reported by ASPCAP can only account for a fraction of the total systematic discrepancy. We interpreted this to mean that \cosha{} systematically predicts slightly cooler and more giant stars than APOGEE. The trend found for \alphaM{} in the \citetalias{Yan2019} discrepancy remains present when comparing to APOGEE, although with a less steep slope. {Since similar trends have been found by other authors when comparing with APOGEE and \citetalias{Yan2019} is largely based on APOGEE estimates, we interpreted this finding as an evidence that this trend is most likely to be originated in a mismatch in the prescriptions used to label APOGEE stars with those used in other studies (including the present).}
    
    \item We predicted the atmospheric properties of the entire (cleaned) MaStar stellar spectra using \cosha{} and characterized the resulting distribution in the parameter space: the most common stars in the library seem to have around $\teff\sim6000\,$K (similar to the Sun's) and being close to the turn-off point. The iron abundance is slightly sub-solar and the $\alpha$-elements abundance relative to iron is slightly above the solar value. This result highlights a deficit of thin disk stars in MaStar, according to \cosha{} predictions. The volume-corrected distributions support these conclusions.

    \item The spatial distribution of the MaStar stars in \alphaM{} vs. \FeH{} for the matched subset of stars in the APOGEE survey reveals an encouraging agreement with the known trends in the $z$ direction of the Galactic plane. The comparison of the full MaStar and APOGEE samples reveals the same trends.
\end{itemize}

In summary we have assigned parameters to the MaStar stellar library using a simple heuristic, non-exhaustive, machine learning approach to predict the atmospheric parameters \teff{}, \logg{}, \FeH{} and \alphaM{} from the optical spectra of $\sim22\,$k unique stars. Our method, dubbed \cosha{}, not only allowed to expand the state-of-the-art empirical libraries in size, but also in dynamical range of the parameter space. The robustness of \cosha{} predictions is clear since without any information about the spatial distribution of the physical properties of the stars in the library, it reproduces the known trends, at least to a \emph{qualitative} level. The version of MaStar presented in this work will grow our ability to analyze resolved and unresolved stellar populations with a precision without precedents.

\acknowledgments

{We thank the anonymous referee for detailed and insightful comments that improved the original manuscript.} We thank CONACYT FC-2016-01-1916 and CB-285080 projects and PAPIIT IN100519 project for support on this study. GB acknowledges financial support from the National Autonomous University of M\'exico (UNAM) through grant DGAPA/PAPIIT IG100319, from CONACyT through grant CB2015-252364 and DGAPA PAPIIT 100622. LC thanks the founding through the PAPIIT IN103820 project. J.B-B acknowledges support from the grant IA-100420 (DGAPA-PAPIIT, UNAM) and funding from the CONACYT grant CF19-39578.

Funding for the Sloan Digital Sky 
Survey IV has been provided by the 
Alfred P. Sloan Foundation, the U.S. 
Department of Energy Office of 
Science, and the Participating 
Institutions. 

SDSS-IV acknowledges support and 
resources from the Center for High 
Performance Computing  at the 
University of Utah. The SDSS 
website is www.sdss.org.

SDSS-IV is managed by the 
Astrophysical Research Consortium 
for the Participating Institutions 
of the SDSS Collaboration including 
the Brazilian Participation Group, 
the Carnegie Institution for Science, 
Carnegie Mellon University, Center for 
Astrophysics | Harvard \& 
Smithsonian, the Chilean Participation 
Group, the French Participation Group, 
Instituto de Astrof\'isica de 
Canarias, The Johns Hopkins 
University, Kavli Institute for the 
Physics and Mathematics of the 
Universe (IPMU) / University of 
Tokyo, the Korean Participation Group, 
Lawrence Berkeley National Laboratory, 
Leibniz Institut f\"ur Astrophysik 
Potsdam (AIP),  Max-Planck-Institut 
f\"ur Astronomie (MPIA Heidelberg), 
Max-Planck-Institut f\"ur 
Astrophysik (MPA Garching), 
Max-Planck-Institut f\"ur 
Extraterrestrische Physik (MPE), 
National Astronomical Observatories of 
China, New Mexico State University, 
New York University, University of 
Notre Dame, Observat\'ario 
Nacional / MCTI, The Ohio State 
University, Pennsylvania State 
University, Shanghai 
Astronomical Observatory, United 
Kingdom Participation Group, 
Universidad Nacional Aut\'onoma 
de M\'exico, University of Arizona, 
University of Colorado Boulder, 
University of Oxford, University of 
Portsmouth, University of Utah, 
University of Virginia, University 
of Washington, University of 
Wisconsin, Vanderbilt University, 
and Yale University.

%



\software{
    python \citep{python1995,python3},
    numpy \citep{numpy2006},
    scipy \citep{scipy2020},
    matplotlib \citep{matplotlib},
    seaborn \citep{seaborn},
    scikit-learn \citep{scikit-learn},
    astroML \citep{astroML},
    astropy \citep{astropy},
    astroquery \citep{astroquery},
    dust-extinction (https://pypi.org/project/dust-extinction/),
    dustmaps \citep{dustmaps}
}



\appendix

\section{Machine Learning: a brief introduction}\label{app:machine-learning}

Machine Learning (ML) can be comprised into a series of statistic and calculus algorithms combined in order to uncover patterns in a data set, without making strong assumptions on the shape of those patterns. The process of uncovering such structures in a given data set is the so called \emph{training} process and its result is the model itself, which can then used to make predictions on new observations. In broad terms, ML algorithms can be classified, regarding the training process, into supervised, semi-supervised, unsupervised and reinforcement learning. In this study we implement the two most common ones: supervised and unsupervised algorithms. We expand on those in the following.

\begin{description}
\item[\textit{Supervised learning}] The main goal of these algorithms is to reveal the shape of the relationship between a given data set arranged in a matrix \fea{} and a set of variables arranged in matrix \lab{}, such that for each record (sample) in \fea{} corresponds a record in \lab{}. The supervised character of these algorithms comes from the fact that the mentioned relationship is learned from a controlled sample, for which the target variable, \lab{} is well-known. Examples of these algorithms are classifications and regressions.
\item[\textit{Unsupervised learning}] In this instance no knowledge is required about the variables \lab{}, instead the learning process consists in finding patterns among the variables in \fea{}. Examples of these algorithms are clustering.
\end{description}



%

\subsection{Precision for GSL}\label{app:gsl-precision}

\begin{figure}
\includegraphics[width=\columnwidth]{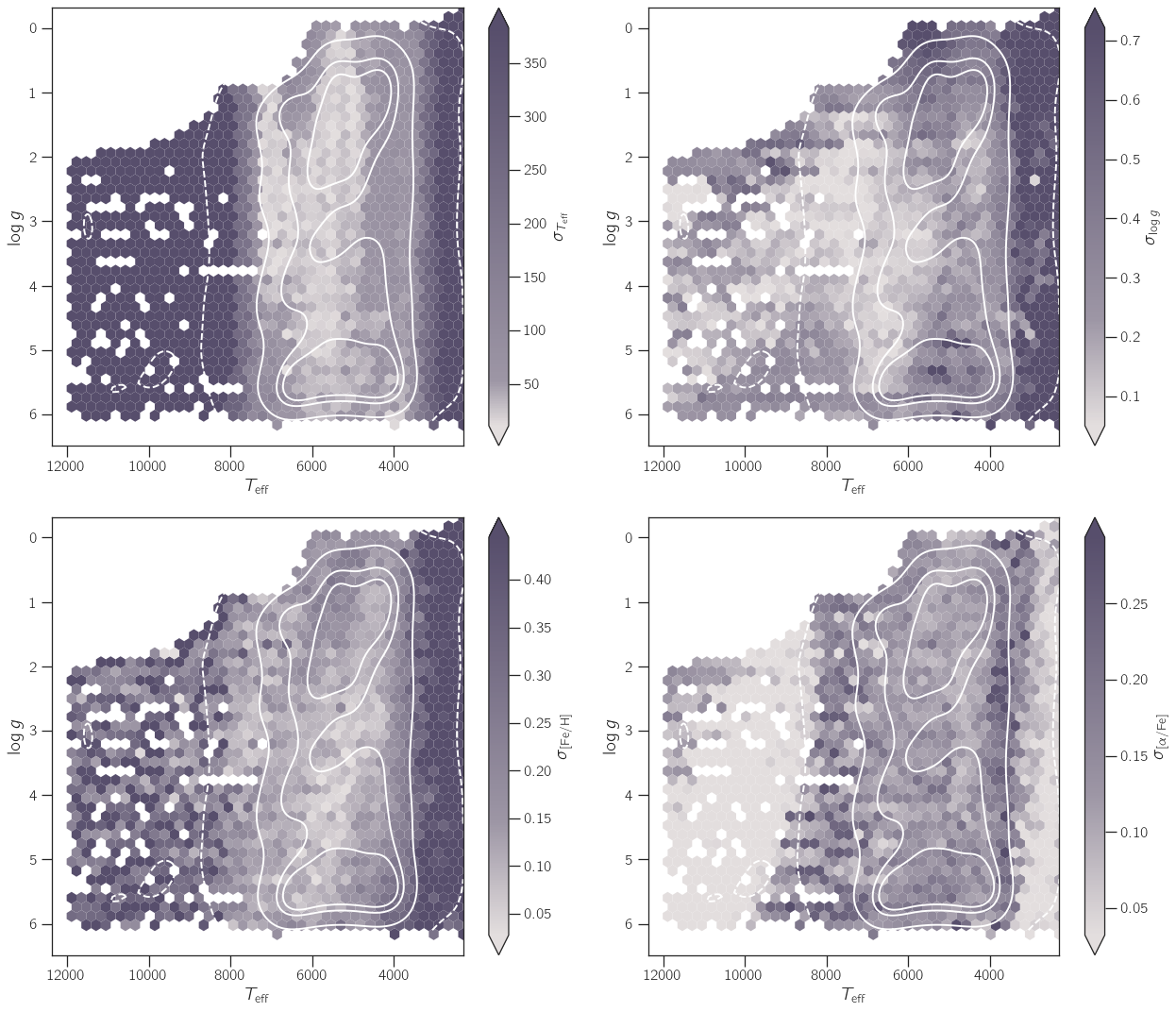}
\caption{Similar to Fig.~\ref{fig:logg-teff-precisions} but for GSL.}\label{fig:logg-teff-precisions-gsl}
\end{figure}

In Fig.~\ref{fig:logg-teff-precisions-gsl} we show the precision of the recovered parameters for GSL using \cosha{}. As expected, the best determination of \teff{} and \FeH{} occurs in the loci of high density in this plane, becoming worse at extreme values of temperature. Interestingly, the \logg{} seems to follow a similar pattern than \teff{}, albeit with a slight improvement towards higher temperatures. The observed pattern for \teff{} and \FeH{} is reversed for \alphaM{} with the best determinations of this parameter being at extreme values of temperature and slightly worsening towards the higher density loci. We remark that, albeit different behaviours in this plot if compared to \ref{fig:logg-teff-precisions}, the level of precision in all parameters remain.

\section{Comparison with other stellar libraries}\label{sec:other-libraries}

\begin{figure}
\includegraphics[width=\columnwidth]{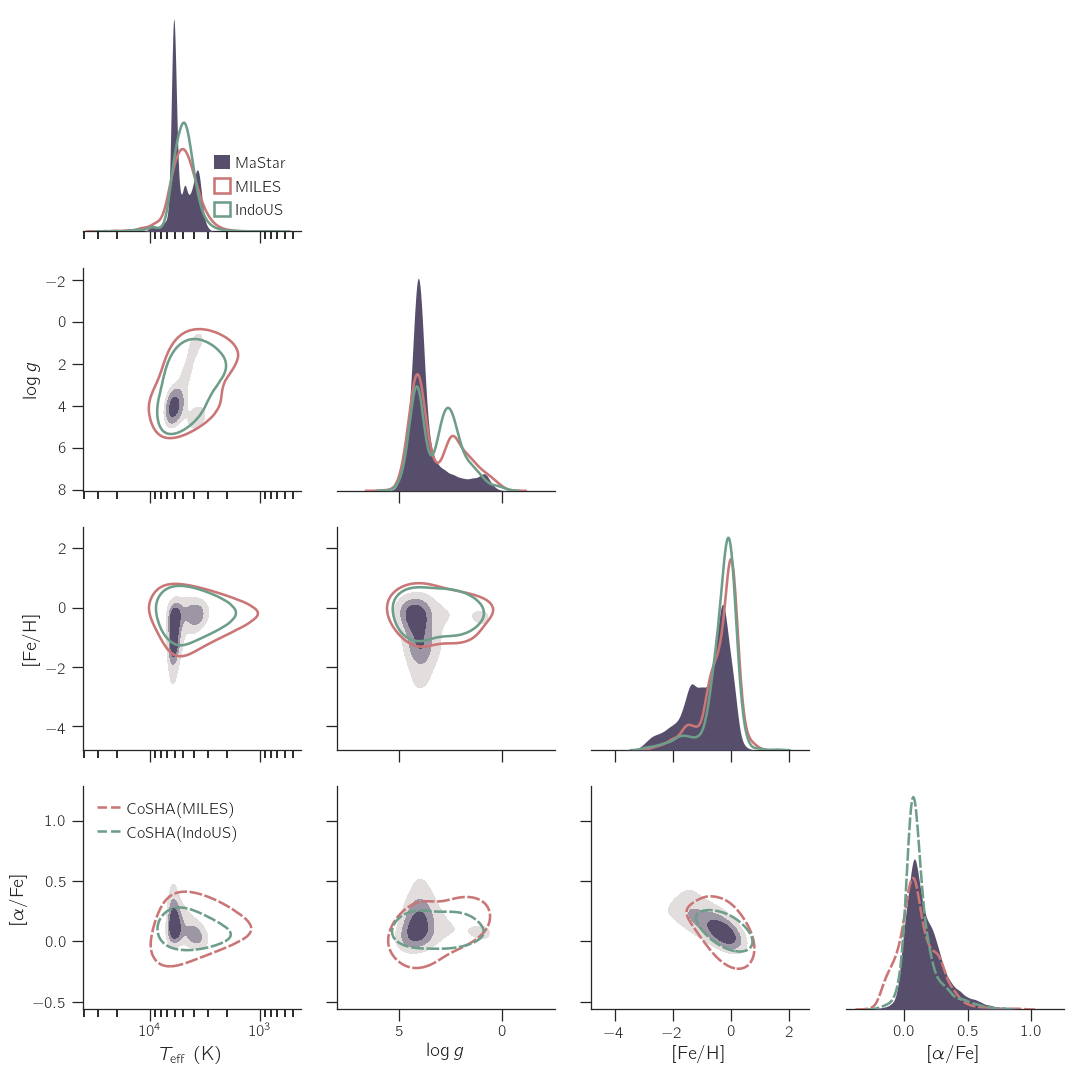}
\caption{Comparison between MILES (pink), IndoUS (green) and MaStar (purple) parameter coverage. MaStar contours enclose the $25\%$, $50\%$ and $75\%$. MILES and IndoUS enclose the $75\%$. MaStar is only superseded by MILES and IndoUS in the high effective temperature side of the parameter space, whilst in \logg{} all libraries have similar coverage and in \FeH{} MaStar extends the lower metallicity extreme by one order of magnitude. We show the \alphaM{} distributions as computed by \cosha{} (dashed lines) for completeness. Again all libraries share a similar coverage in this parameter.}\label{fig:libraries-comparison}
\end{figure}

In this section we explore to what extend the MaStar library as analized by \cosha{} improves the sampling of the parameter space upon previous empirical libraries. This is not meant to be an exhaustive exploration, but merely a comparison with widely used stellar libraries. For this purpose we choose the IndoUS \citep{Valdes2004} and MILES \citep{Sanchez-Blazquez2006, Cenarro2007} stellar libraries. Since we both these libraries are freely available to download, we analyze the corresponding spectra using \cosha{} and then compared the resulting parameter distribution with the one distributed with those libraries. There is a caveat though: the wavelength range coverage. As a matter of fact, MaStar, MILES and IndoUS do not have the same coverage of the optical wavelength range, nor the same sampling and resolution. We use the same pre-processing procedure described in \S~\ref{sec:preprocessing}. We fill in the missing pixels in MILES ($\lambda\sim7,500$~---~$10,000\,$\AA) and in IndoUS ($\lambda\sim9,500$~---~$10,000\,$\AA) using as a reference the MaStar spectra. We compared the performance of \cosha{} with and without filling in the missing pixels in MILES and IndoUS and found a considerable improvement in the parameter space after extending the spectral range in those libraries. Another difference is that both this libraries have no publicly available \alphaM{} estimate \citep[however see][in the case of MILES]{Knowles2021, Coelho2020}, hence we limit our consistency tests to \teff{}, \logg{} and \FeH{} only.

In Fig.~\ref{fig:libraries-comparison} we show the parameter space coverage of the MILES, IndoUS and MaStar libraries, for comparison purposes. MILES, IndoUS and MaStar lowest contour enclose $75\%$ of the density distribution. For completeness we show the \alphaM{} distributions of MILES and IndoUS computed using \cosha{} (dashed distributions). Clearly MILES and IndoUS span a wider range in \teff{}, reaching to $\sim30,000\,$K as opposed to $12,000\,$K for MaStar. In \logg{} all libraries have similar coverage, with the MaStar extending the limits only marginally. In \FeH{} MaStar extends the lower end of the distribution, reaching down to $\sim-4\,$dex as opposed to $\sim-3\,$dex for MILES and IndoUS. We recall that the limitations in the MaStar parameter coverage is likely to have its origin in the training set and can be lifted by including hotter stars. Albeit such weaknesses, the version of the MaStar presented throughout this study has an important advantage: since MaStar has over one order of magnitude more stars than these stellar libraries, it provides the best sampling of the parameter to date by any empirical library.

\newpage
\bibliographystyle{aasjournal}
\bibliography{mastar,external} 

\end{document}